\pdfoutput=1
\documentclass[manuscript,nonacm]{acmart}

\usepackage{tikz}
\newcommand{\circled}[1]{
  \tikz[baseline=(char.base)] \node[shape=circle,draw,inner sep=-0.5pt] (char) {\strut #1};}

\usepackage{xcolor}
\usepackage[colorinlistoftodos,prependcaption,textsize=tiny]{todonotes}

\usepackage{soul}
\usepackage{longtable}

\usepackage{xspace}  
\usepackage{algorithm}
\usepackage{algorithmic}
\usepackage{microtype}
\usepackage{colortbl}
\usepackage{enumitem}
\usepackage[most]{tcolorbox}

\definecolor{customPurple}{HTML}{6C63FF}
\newcommand{\method}{\textsf{CoPrompter}\xspace}

\newif\ifcomments

\commentstrue
\ifcomments
    \providecommand{\ishika}[2][]{{\protect\color{magenta}{[\textbf{Ishika}:\textbf{#1} #2]}}}
    \providecommand{\shreeya}[2][]{{\protect\color{red}{[\textbf{Shreeya}:\textbf{#1} #2]}}}
    \providecommand{\manushree}[2][]{{\protect\color{violet}{[\textbf{Manushree}:\textbf{#1} #2]}}}
    \providecommand{\simra}[2][]{{\protect\color{orange}{[\textbf{Simra}:\textbf{#1} #2]}}}

\else
    \providecommand{\ishika}[2][]{}
     \providecommand{\shreeya}[2][]{}
     \providecommand{\manushree}[2][]{}
     \providecommand{\simra}[2][]{}
\fi

\AtBeginDocument{%
  }
    
\makeatletter
\def\@copyrightspace{\relax}
\makeatother
\author{Ishika Joshi\textsuperscript{1,2 *$\psi$}, Simra Shahid\textsuperscript{1,2 *$\psi$}, Shreeya Venneti\textsuperscript{4*\textdagger}, Manushree Vasu\textsuperscript{5*\textdagger}, Yantao Zheng\textsuperscript{1}, Yunyao Li\textsuperscript{1}, Balaji Krishnamurthy\textsuperscript{1,2}, Gromit Yeuk-Yin Chan\textsuperscript{1,2}}

\authornote{\textbf{\textsuperscript{*}First authorship. \textsuperscript{$\psi$}\ Principal Investigators. \textsuperscript{\textdagger}\ Work done in Internship with Adobe Media and Data Science Research.}}
\authornote{Correspondence to Ishika Joshi (ijoshi@adobe.com), Simra Shahid (sshahid@adobe.com).}

\affiliation{
  {\textsuperscript{1}Adobe,
  \textsuperscript{2}Media and Data Science Research,
  \textsuperscript{3}Adobe Research, 
  \textsuperscript{4}International Institute of Information Technology Bangalore, 
  \textsuperscript{5}Georgia Institute of Technology}\country{\textcolor{white}{U}}
}

\title{CoPrompter: User-Centric Evaluation of LLM Instruction Alignment for Improved Prompt Engineering}

\begin{document}

\begin{teaserfigure}
    \centering
    \includegraphics[width=1\textwidth]{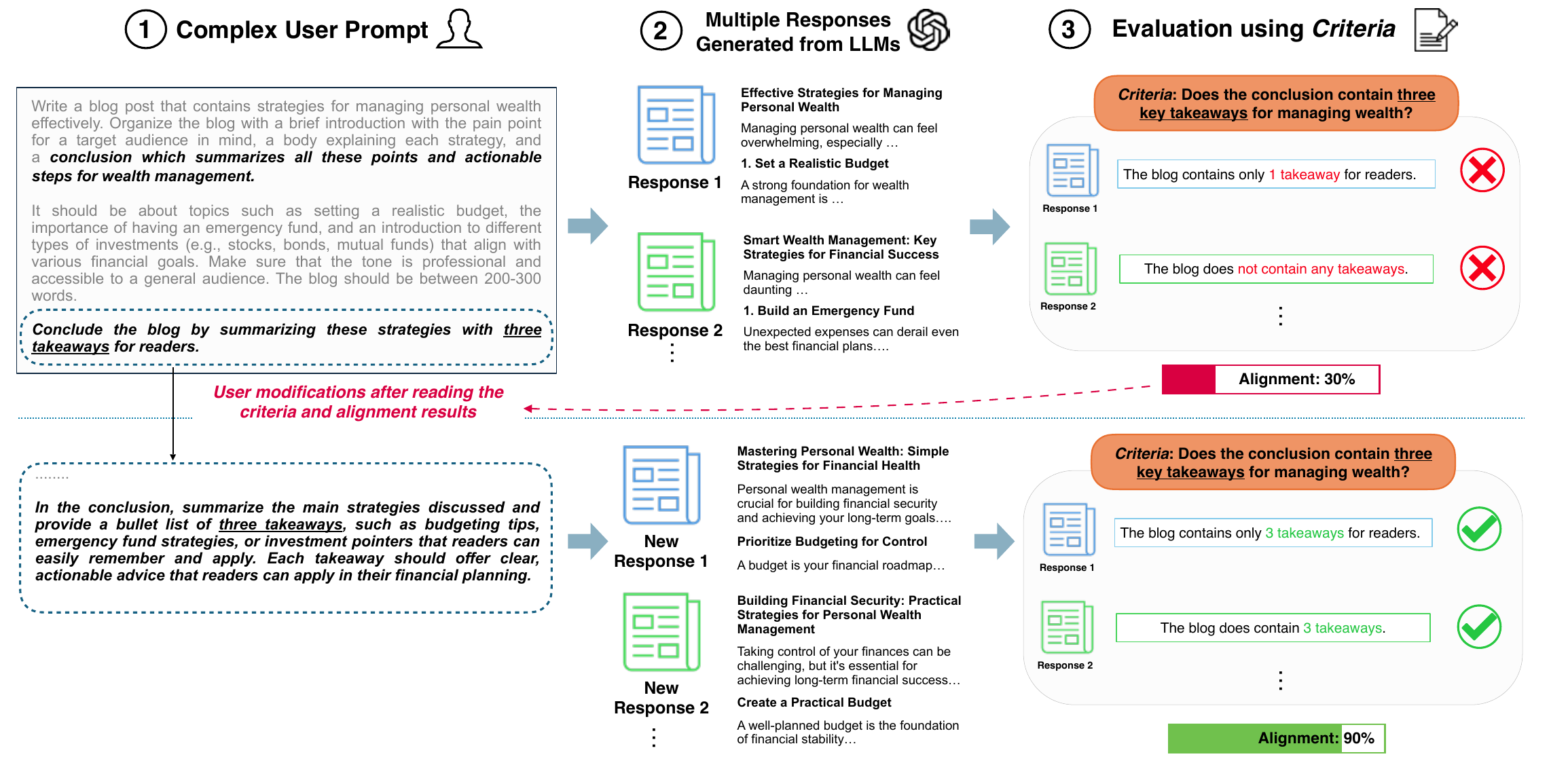} 
    \caption{Overview of \method framework: Given a \textcircled{\raisebox{-0.9pt}{1}} complex prompt from the user and \textcircled{\raisebox{-0.9pt}{2}} multiple responses generated by LLMs, \textcircled{\raisebox{-0.9pt}{3}} \method generates a set of \textit{criteria} to assess the responses and produce an alignment score. Users could iteratively refine and reflect their prompts to result in a prompt that achieves better alignment.}
    \label{fig:abstract_figure}
\end{teaserfigure}

\begin{abstract}
    Ensuring large language models' (LLMs) responses align with prompt instructions is crucial for application development. Based on our formative study with industry professionals, the alignment requires heavy human involvement and tedious trial-and-error especially when there are many instructions in the prompt. To address these challenges, we introduce \method, a framework that identifies misalignment based on assessing multiple LLM responses with criteria. It proposes a method to generate \textit{evaluation criteria questions} derived directly from prompt requirements and an interface to turn these questions into a user-editable checklist. Our user study with industry prompt engineers shows that \method improves the ability to identify and refine instruction alignment with prompt requirements over traditional methods, helps them understand where and how frequently models fail to follow user's prompt requirements, and helps in clarifying their own requirements, giving them greater control over the response evaluation process. We also present the design lessons to underscore our system's potential to streamline the prompt engineering process.

\end{abstract}

\maketitle

\section{Introduction}

Large language models (LLMs) have become ubiquitous in a wide range of applications, finding use across billion-dollar industries for tasks ranging from basic operations to critical applications~\cite{1_Introduction, 2_Introduction, 3_Introduction, 4_Introduction, 5_Introduction}. Prompting has emerged as the primary method for developers to interact with LLMs, enabling rich, Generative AI-driven experiences. 

Given that many critical products and services rely on LLMs and prompts, ensuring that the generated contents align with user expectations is crucial. 

Recent works have explored emerging themes of \textit{misalignment} between prompt engineers and model responses. Drawing from the definition provided by Terry et al.~\cite{InteractiveAIAlignment}, we define \textit{misalignment} as ``LLMs generating outcomes that deviate from the goals or expectations of the user, often leading to unintended or undesirable effects''. Some misalignments that have been explored in literature include instruction overlooking and misinterpretation \cite{WhyJohnnyCantPrompt, ChainOfThoughtPrompting, 9_Introduction_MoreLLMIssues, 24_Intro_Risks2, 25_Intro_Risks3, 26_Intro_Risks4}  hallucinations~\cite{6_Introduction_Hallucinations, 12_Introduction_Hallucination2, 13_Introduction_Hallucination3, 14_Introduction_Hallucination4, 15_Introduction_Hallucination5}, generation of harmful~\cite{7_Introduction_HarmfulContent, 16_Introduction_Harmful2, 17_Introduction_Harmful3, 18_Introduction_Harmful4, 19_Introduction_Harmful5} or biased~\cite{8_Introduction_BiasedResponses, 20_Introduction_Bias2, 21_Introduction_Bias3, 22_Introduction_Bias4, 23_Introduction_Bias5} contents, etc. This often stems from the inherent sensitivity of LLMs to the words within the prompt and the overall structure of prompts, where even minor changes can lead to drastically different outputs ~\cite{10_Introduction_PromptFormatSensitivity, 27_Intro_PromptSensitivity, 28_Intro_PromptSensitivity, 29_Intro_PromptSensitivity}. 

Despite efforts being made towards improving the instruction-following ability of LLMs by training them on instruction datasets~\cite{11_Introduction_InstructionDatasetsFinetuning, 1_RelatedWork, 8_RelatedWork, 9_RelatedWork}, existing approaches often fall short of being consistently aligned with all the instructions in long-form prompts with numerous instructions  (Figure~\ref{fig:abstract_figure}). These prompts tend to experience greater misalignment not only due to the complexity of multiple instructions but also the variations in instruction writing styles that causes the model to overlook or misinterpret the prompt engineer's intended outcomes.

To better understand the practical challenges faced by prompt engineers, we conducted a formative survey with 28 professionals working across various domain-specific industry applications (Section \ref{sec:formative_study}). Based on the survey, the misalignment issues such as \textit{overlooked instructions, inconsistent responses, instruction misinterpretations}, and \textit{incorrect assumptions} are frequent, particularly when dealing with complex prompts containing five or more instructions. The current workflow for prompt engineering is thus \textit{time-consuming} and \textit{tedious}, often requiring over ten iterations and \textit{manual inspection} of responses in a largely \textit{trial-and-error} process.

In response to these challenges, we introduce \method, a novel tool that helps prompt engineers systematically identify and address areas of misalignment between multiple LLM outputs and their requirements. \method achieves this by first breaking down user requirements into atomic instructions, each transformed into criteria questions. With the criteria and LLM outputs, \method generates detailed misalignment reports at the instruction level. This granular approach allows users to quickly pinpoint where misalignments occur and identify which criteria fail most frequently, offering a systematic way to prioritize prompt refinements. As a \textit{user-in-the-loop} tool, \method provides prompt engineers with control over the evaluation process, allowing them to customize criteria to better reflect specific requirements and adapt to evolving needs.


This work provides the following contributions:
\begin{itemize}
    \item We introduce \method, a novel user-in-loop system that helps prompt engineers align complex prompts with their requirements through systematic evaluation and customizable criteria control.

    \item A technical pipeline that generates detailed misalignment reports, allowing prompt engineers to pinpoint specific areas in need of improvement and prioritize prompt adjustments.

     \item We present findings from a formative study with 28 prompt engineers, contributing to the understanding of the challenges in crafting complex prompts and the common misalignments observed with LLM responses.

    \item We conduct a comprehensive user evaluation with 8 industrial prompt engineers, demonstrating that \method effectively identifies misalignments, helps in prompt refinement, and adapts to evolving requirements, significantly improving upon traditional manual methods. 
    
    \item A focused user evaluation with high System Usability Scale (SUS) scores further reflects user confidence in its seamless integration into prompt improvement workflows.

\end{itemize}

\section{Related Work}

\subsection{Challenges in Aligning LLM Outputs with Prompt Instructions}

Misalignments between LLM outputs and user instructions are a well-documented issue. Among the various alignment goals~\cite{20_AlignmentType, 21_AlignmentType}, our research focuses on the LLM's ability to understand and follow long-context instructions. A model is considered \textit{misaligned} when it deviates from expected outcomes, producing ambiguous, sub-optimal, or even harmful results~\cite{4_RelatedWork, 5_RelatedWork, 6_RelatedWork}. These misalignments can occur when the model overlooks or misunderstands instructions, hallucinates content, or produces inconsistent responses across multiple interactions.

Previous studies have highlighted the difficulty in ensuring LLMs consistently follow prompts, especially those containing multiple instructions~\cite{7_RelatedWork, 8_RelatedWork, 9_RelatedWork, 10_RelatedWork, 14_RelatedWork_Dolomites, 15_RelatedWork_FollowBench, 16_RelatedWork_InfoBench, 17_RelatedWork_Suri}. While these works propose to improve such alignment issues using approaches like supervised fine-tuning (SFT), preference tuning (e.g. DPO~\cite{12_RelatedWork_SFT}, RLHF \cite{13_RelatedWork_PreferenceTuning}, ORPO~\cite{17_RelatedWork_Suri}) and instruction finetuning on these datasets have helped improve this alignment issues, there are more recent works that continue to report more alignment issues that surface. The HCI scholarship has explored these challenges and attributed them to discrepancies in Human AI communication \cite{28_RelatedWork_BeyondPrompts}. While alignment has been studied from a standard benchmark perspective, researchers in HCI communities have highlighted the dynamic nature of alignment. Shen et al. emphasized that alignment is a bi-directional concept, requiring both the alignment of AI systems to human factors and the adaptation of humans to the capabilities of AI~\cite{TowardsBidirectionalHumanAIAlignment,TowardsHumanCenteredAICoCreation}. Moreover, unpredictability, lack of transparency, value misalignment, and inherent complexity in understanding response behaviors while interacting with LLMs have been identified as common causes of misalignment \cite{Dung2023, ReexaminingWhetherWhyHow, ReexaminingWhetherWhyHow}. Addressing misalignments becomes a complex process for prompt engineers who often try to optimize prompts while navigating these challenges. This highlights the need to build systems that bridge the gap between LLMs and prompt engineers by easing the process of identifying misalignments systematically, assisting prompt engineers in discovering the cause of misalignment and addressing it for streamlined prompt improvement.

\subsection{Systematic Assessment of LLM Outputs}

To systematically assess and improve LLM outputs, researchers have explored the use of evaluation criteria and LLM-as-a-Judge. A criterion serves as a structured measure for evaluating whether an LLM response meets specified conditions. It typically comprises a clear context of the condition, examples illustrating both ideal and non-ideal outputs, and a scoring system or rubric to standardize assessment\cite{18_RelatedWork_AutoCalibrate}. Prior work has employed criteria to evaluate aspects such as coherence, naturalness, and grounding in summarization tasks \cite{18_RelatedWork_AutoCalibrate, 19_LLMReferenceFreeQuality, 22_RelatedWork_HDEval}. 

\noindent Prior work has also introduced various tools \cite{27_RelatedWork_PromptMaker, 28_RelatedWork_BeyondPrompts, 23_RelatedWork_ChainForge, 24_RelatedWork_SPADE, 25_RelatedWork_EvalLM, shankar2024validates},
PromptsRoyale\footnote{\url{https://www.promptsroyale.com/}}, promptfoo\footnote{\url{https://www.promptfoo.dev/}} to help evaluate model responses and support different needs of prompt engineers. ChainForge \cite{23_RelatedWork_ChainForge} enables users to generate prompts, test them across multiple LLMs, and compare outputs side by side. It supports custom evaluation metrics and helps users iterate on prompt design. SPADE \cite{24_RelatedWork_SPADE} proposes automatically generating criteria based on changes in prompt versions, helping users analyze what improves response quality. It assists in generating Python assertion functions for LLM outputs, aligning evaluations with user expectations. EvalLM \cite{25_RelatedWork_EvalLM} assists changes in prompt design into a collaborative process, where the LLM acts as both an evaluation assistant and criteria reviewer, identifying potential improvements in criteria quality.
EvalGen \cite{shankar2024validates} builds upon ChainForge and enables users to create evaluation pipelines that align with their preferences facilitating user-specified and automated criteria checks to assess LLM outputs automatically. Further, it allows users to assign ground truth to each criterion which they coin as candidate assertions, and generate reports on criteria alignment. 

While these systems serve the purpose of enabling user-aligned evaluation, however, they do not assist prompt engineers in identifying causes of overall misalignment between user expectations and generated responses. Prompt Engineers still struggle with identifying all the points of misalignment in their prompts through manual inspection of responses and using arbitrary techniques to make modifications and improve prompt alignment.
Thus, we address this challenge by expanding on the human-in-the-loop evaluation framework introduced by Shankar et al.~\cite{shankar2024validates}, using criteria derived from atomic instructions within prompts to evaluate LLM response alignment with user-defined requirements. We build upon the framework in three key areas: (i) Instead of Python assertions, we automatically create criteria derived from the entire user prompt input, allowing for more complex criteria that serve as a checklist of sub-instructions within the prompt. The criteria generated also inform users of the nature of these criteria for eg. identifying any subjectivity and can be interpreted differently. (ii) Next, we provide support for customizable criteria evaluation. This enables prompt engineers to evaluate their changing requirements at any point in the form of editable criteria questions. (iii) Finally, our work differs by emphasizing providing detailed reports of LLM response alignment to user specified guidelines/requirements, which helps users to iterate on their prompts more systematically.

\section{Formative Study}\label{sec:formative_study}

\noindent The motivation \method stems from findings from a formative study involving a survey of 28 industry prompt engineers, who have leveraged LLMs to develop domain-specific applications across content generation, writing assistance, chatbot development, summarization, query rewriting, and email drafting, among others. The study survey was circulated across Slack channels and email threads of prompt engineers across an IT company. To better understand their prompt engineering process and the issues they encounter, we asked the participants detailed questions (Appendix \ref{sec:formative-study-questionaire}). Specifically, we asked questions around three key areas:

\begin{itemize}
    \item \textbf{Instruction Styles}: Here, we examined how prompt engineers structure their prompts, focusing on whether they use concise lists, detailed step-by-step guides, or other styles. We further classified instruction types, such as context-setting, format specifications, style or task-specific instructions.
    Notably, \textbf{64\% of participants reported that their typical prompts contain 5-10 instructions, while over 55\% use even more complex prompts with more than 10 instructions}. These prompts often contain task-specific instructions, formatting guidelines, and additional instructions to provide context to guide the LLM’s response. Most participants structure their prompts concisely and use either point-based lists or use examples to clarify instructions, while a small number of users write prompts in paragraph format. 

    \item \textbf{Challenges of Instruction Following Misalignment}: We asked the participants if they found any discrepancies between LLM responses with prompt instructions. Participants were asked to rate  the level of alignment of generated responses on a scale of 1 to 5 for their first few prompts. 
    Most participants rated the initial alignment of LLM responses as a 3 or 4 on a 5-point scale (where 5 indicates perfect alignment), suggesting that while responses generally aligned with the prompt’s intent, they were not always precise. However, when asked about the number of attempts it takes to reach a prompt that generates desirable responses, the majority of the participants expressed that it took 10+ attempts and evaluation of 40+ responses to reach an optimized prompt. 
    We also asked them to describe any common types of misalignment observed between their instructions and the LLM’s outputs. From their responses, a key challenge identified was the different types of \textbf{misalignment} including \textbf{overlooked or ignored instructions, hallucinated content, misinterpreted details, and inconsistent responses across multiple interactions}. The complex prompts with detailed task or formatting instructions were prone to misalignment, even with clear and elaborate guidance provided.

    \item \textbf{Strategies for Navigating Instruction Following Misalignment}: To address misalignment, participants reported using several corrective strategies, such as identifying misinterpreted instructions and improving them by providing additional context. A common approach involved breaking down complex instructions into smaller steps and rearranging instructions for clarity. However, participants described this \textbf{process as highly time-consuming and a trial-error process}. Nearly all participants noted that achieving desired alignment typically \textbf{required over 10 iterations, with some complex prompts taking even longer}. This iterative, trial-and-error approach underscores the need for a more structured, systematic solution for prompt improvements.

\end{itemize}

\section{Design Goals}
\label{sec:design_goals}

In designing \method, we aim to address the challenges that prompt engineers face when aligning complex prompts with their requirements. 
Based on our formative study and literature review, we identified three design goals to guide the development of \method:

\vspace{-1mm}
\begin{itemize} 

\item \textbf{(DG1)}: \textsf{Help prompt engineers in  \textbf{aligning complex prompts with their requirements} through systematic evaluation.}

In the formative study, participants often used complex prompts containing more than five instructions, and observed different types of misalignment (overlooked instructions or misinterpretations). However, they lacked a systematic method to evaluate whether each instruction in the prompt was being followed by the LLM.

\method addresses this by converting user prompt requirements, referred to as \textit{guidelines}, into \textit{evaluation criteria questions}. To reduce overloaded information and ensure clarity, \method decomposes these guidelines into atomic instructions, each representing a single requirement. This decomposition makes it easier to identify and address specific areas of misalignment.

Additionally, \method applies appropriate evaluation methods tailored to each criterion, guiding the evaluation process with appropriate methods for each type of criterion. This systematic evaluation helps in aligning complex prompts with the user's requirements, providing clarity on which parts of the prompt need refinement.

\item \textbf{(DG2)}: \textsf{Help prompt engineers \textbf{refine their prompts} using insights from misalignment reports between model responses and instructions.} 

For every generated prompt response, \method collects all the misalignment and alignment observed at instruction level (for each criteria), and provides prompt engineers with a detailed alignment report. This report consolidates areas where requirements were not met and shows which criteria fail most frequently, helping prompt engineers prioritize changes. By pinpointing specific instructions that require attention, \method guides users in refining their prompts to improve alignment with their intended requirements.

\item \textbf{(DG3)}: \textsf{Provide support for \textbf{evolving requirements} and support continuous prompt refinement.}

Prompt engineering is an iterative process, where user requirements may shift as prompt engineers gain new insights or refine their understanding of the task. \method addresses this by allowing continuous updates to evaluation criteria, so prompt engineers can add, delete, or modify criteria questions as needed. Furthermore, \method highlights perceived subjectivity in criteria, alerting users to criteria that may need further clarification. This user-in-the-loop approach ensures that \method remains adaptable to changing requirements, allowing prompt engineers to keep their evaluation criteria aligned with the evolving needs of their task.

\end{itemize}

\section{\method: System Design}

\begin{figure}[tb!]
    \centering
   \includegraphics[width =\textwidth]{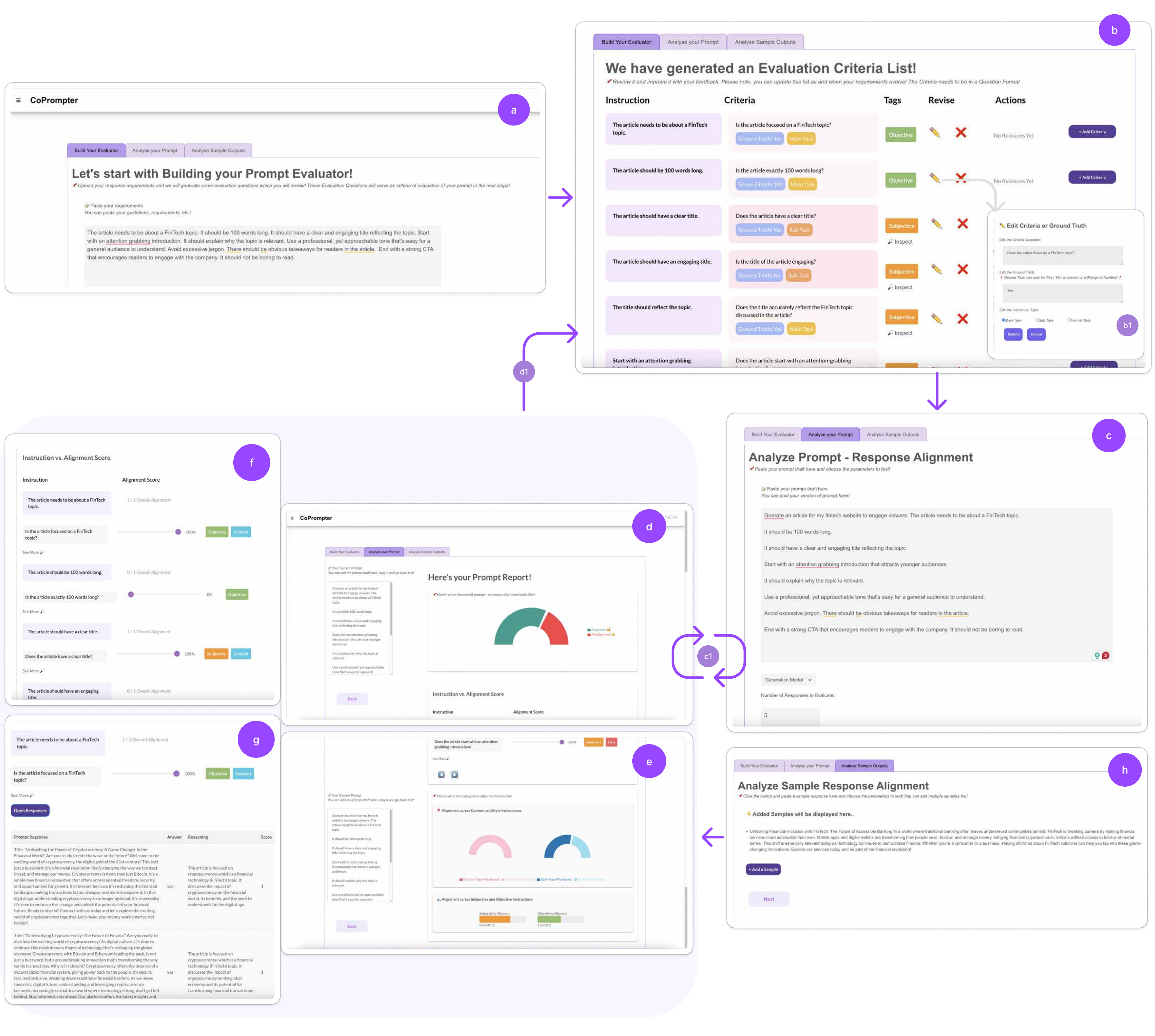} 
    \caption{
    (a) Users enter prompt requirements in the 'Build Your Evaluator' tab. (b) Atomic instructions and criteria questions are extracted for evaluating responses, with (b1) options to modify them before saving. (c) In the 'Analyse Your Prompt' tab, users input a prompt, select a model, and choose the number of responses for evaluation. (d) Responses are evaluated based on the criteria, and alignment scores are displayed in the Alignment Report Card. (e) Users can view detailed scores, generated responses, \method's score, and rationale, and adjust the prompt accordingly. (f) \method categorizes alignment by content, style, and instruction type. (g) The prompt can be iteratively improved and retested. (h) Sample responses can also be tested against the evaluation criteria.
    }
    \label{fig:ui-flow}
\end{figure}

In this section, we describe how \method system meets the design goals (Section~\ref{sec:design_goals}). Section \ref{sec:system_frontend} outlines the User Interaction Workflow (Figure~\ref{fig:ui-flow}). Section~\ref{sec:system_backend} covers the backend implementation details (Figure~\ref{fig:system_architecture}). 

\subsection{User Interaction Workflow}\label{sec:system_frontend}

\method facilitates an iterative process where users define, refine, and evaluate prompt responses through a structured workflow. As illustrated in Figure \ref{fig:ui-flow} (with zoomed-in images in the Appendix), this interactive interface enables users to review and modify evaluation criteria, generate prompt responses using LLMs, and assess these responses for alignment with their specified requirements.

\subsubsection{\textbf{Build Your Evaluator:}}
\label{sec: build_your_eval}
In the `Build Your Evaluator' stage (a,b,b1 in Figure \ref{fig:ui-flow}, users set up the evaluation framework based on their specific task requirements. Users begin by providing prompt requirements or guidelines, outlining what they expect in generated responses. This can be a prompt draft or guidelines that include the instructions that the user wants to evaluate in the generated prompt responses. This stage is supported by two backend modules.

\noindent \circled{1} \textbf{Evaluation Criteria Generation Module}. This module breaks down these high-level guidelines into detailed evaluation questions, known as criteria questions (CQs) which help in comparing the alignment between the generated prompt responses and user’s intended requirements.

\begin{figure}[t]
    \centering
    \includegraphics[width=\textwidth]{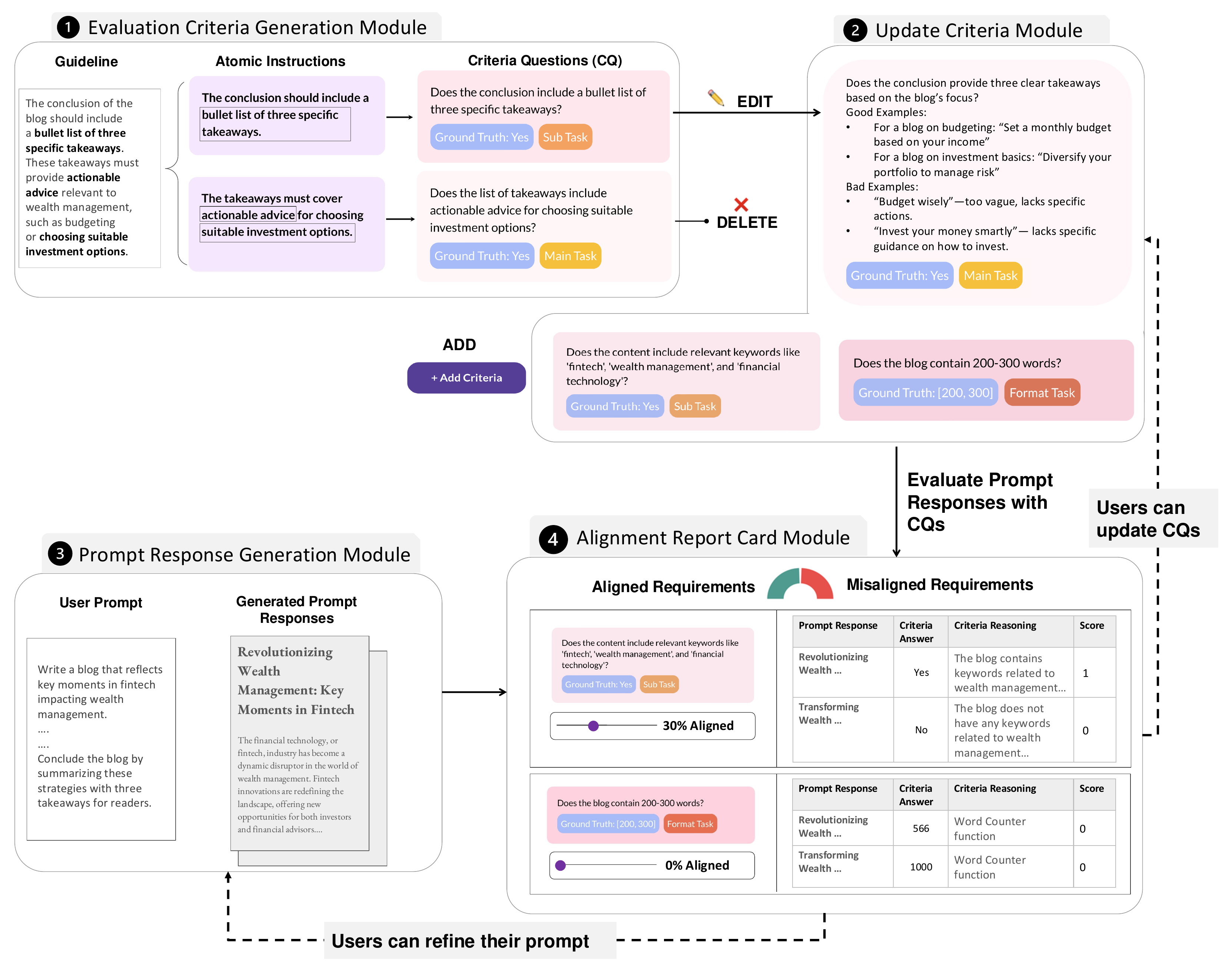}
    
    \caption{The \method system workflow 
    begins with (1) the Evaluation Criteria Generation Module, where user prompt requirements (guidelines) are broken down into atomic instructions, which are then formulated as criteria questions (CQs) with ground truth labels and metadata tags (e.g., main task, subtask, or format-related). Users can adjust these criteria in (2) the Update Criteria Module by editing, deleting, or adding new CQs. Next, in (3) the Prompt Response Generation Module, users input their prompt, select an LLM, and generate responses. The responses are evaluated against the defined criteria, with results displayed in (4) the Alignment Report Card Module. This report shows aligned and misaligned requirements, allowing users to explore misalignments in detail. Based on the feedback, users can refine their prompt or criteria, iterating as needed to improve alignment.
        }
    \label{fig:system_architecture}
\end{figure}

\textit{For example}, as shown in Figure \ref{fig:system_architecture}, we assume the user persona of a content creator who wants to use LLMs to generate finance blogs. The user may provide a guideline that checks if the blog’s conclusion includes a `\textit{bullet list of specific takeaways with actionable advice for investment options}'. This instruction in the guideline is decomposed into two \textit{atomic} instructions: `\textit{bullet list of three specific takeaways}' and `\textit{actionable advice for investment options}'. These atomic instructions are then converted into criteria questions: 

\begin{enumerate}
    \item \begin{quote}
    \textbf{Extracted Atomic Instruction:} \textit{The conclusion should include a bullet list of three specific takeaways}.
    \\
    \textbf{Corresponding Criteria Question:} Does the conclusion include a bullet list of three specific takeaways? \\ \textbf{Ground Truth:} Yes
\end{quote}
\item \begin{quote}
    \textbf{Extracted Atomic Instruction:} \textit{The takeaways must cover actionable advice for choosing suitable investment options}.
    \\
    \textbf{Corresponding Criteria Question:} Does the list of takeaways include actionable advice for choosing suitable investment options? \\ \textbf{Ground Truth:} Yes
\end{quote}

\end{enumerate}

For each criteria question we also generate ground truth labels, and priority tags (e.g., main task, subtask, or format-related) that help in structured evaluation. \method also provides metadata tags, such as content vs. style categorization and question subjectivity. This ground truth is crucial for computing alignment and is compared with LLM-generated ground truth for each prompt response. Initially, users may only have a general understanding of their task requirements, but they can refine their criteria as their needs evolve during the evaluation setup.

\noindent \circled{2} \textbf{Update Criteria Module}: After generating initial criteria questions, users can refine them by editing, deleting, or adding new criteria. This user-in-the-loop module offers flexibility and control, allowing updates to questions, ground truth, or priority tags as evaluation goals become clearer. Figure \ref{fig:system_architecture} outlines examples of specific edits and other ways to refine criteria.

\begin{itemize}
    \item \textsf{Edit}: 
    Users can edit a criteria question to make it more specific, they can include definitions or examples as shown in Figure \ref{fig:system_architecture}.
   
Users can view categorical tags for each criteria question to pinpoint subjectivity areas needing clarification. The formative study indicated that subjective instructions often result in misalignment due to differing interpretations between the prompt engineer and the model. To address this, \method classifies criteria as either 'Subjective' or 'Objective,' helping users identify those that require additional context. For subjective criteria like 'engaging tone' or 'concise summary,' \method provides multiple interpretations with examples—for instance, a 'concise summary' may mean 2-3 lines in one context but 10-15 in another. These interpretations allow users to add clarifications to align with their expectations. All changes are tracked and versioned to ensure only the latest version is used during evaluation.

    \item \textsf{Delete}: If the user feels that a criteria is not important for evaluation or is redundant by another criterion, users can choose to delete it.
    \item \textsf{Add New Criteria}: Users can add new criteria for an existing guideline instruction or a completely new requirement      
    to capture additional requirements. They can add criteria which are of the following types:
    \begin{itemize}
        \item Measurable: If users want shorter responses, they might add a criterion specifying a word count, either as an exact number (e.g., =200 words) or an acceptable range (e.g., [200, 300] words).
        \item Descriptive: Users may want to check for specific type of keywords or if the conclusion contains bullet points, with a simple yes/no ground truth.
        \item Layered Measurable allows users to set complex evaluation criteria that first identify a specific section (e.g., the conclusion) and then apply a measurable check, such as, \textit{Does the conclusion have fewer than 50 words?} This approach enables targeted evaluation of specific parts of a prompt response.
    \end{itemize}
    
\end{itemize}

The \textbf{Build Your Evaluator} stage offers a flexible framework for evaluating alignment by providing user-in-the-loop control over question content, ground truth, and tags. This approach allows users to continuously refine their criteria to match evolving task requirements. Users can then \textit{save their criteria} and proceed to the next stage for evaluating their prompt.

\subsubsection{\textbf{Analyse Your Prompt:}}
In this stage, users evaluate how well generated responses align with the criteria defined in the previous stage (c,c1,d,e,f,g in Figure \ref{fig:ui-flow}). This stage is supported by two backend modules:

\noindent \circled{3} \textbf{Prompt Response Generation Module}. Within the `Analyse Your Prompt' tab, users input their prompt draft, select a language model, and specify the number of responses they wish to generate. This module processes these inputs and generates multiple responses specified by the user using the LLM of their choice. These responses are independently evaluated against each saved criteria. 

\noindent \circled{4} \textbf{Alignment Report Card Module}.

After the responses are generated, the Alignment Report Card Module evaluates each response against the saved criteria questions, providing a structured report on alignment that shows where responses meet or dont align with user requirements (criteria). This report is presented in an interactive \textbf{Prompt Alignment Report} that displays alignment scores in different dimensions: overall, by category, and for each specific criterion. 

To start, the report shows a pie chart summarizing the number of aligned and misaligned criteria across all responses, giving users a quick sense of how well the responses match their expectations. Next, users can explore a detailed breakdown for each Atomic Instruction and its matching Criteria Question. This breakdown shows exactly which requirements are being followed and where there are gaps. Users can click `See More' to view each response, its alignment score, and an explanation from \method, providing transparency in how each response is scored.

For example, as shown in Figure \ref{fig:system_architecture}, the report evaluates two user-defined criteria: 

\begin{itemize}
    \item Checking for Keywords: This criterion checks for the presence of keywords like 'wealth management' in the blog, with only 30\% of generated blogs meeting this criterion. On the right side, a table displays results similar to the 'See More' option in the UI. In this example, the first blog includes the keywords, receiving a \textit{yes} in the Criteria Answer column and a score of 1, indicating alignment with user-approved ground truth. The second blog lacks these keywords, resulting in a score of 0 due to a mismatch with ground truth.
    
    \item Word Count: 
This criterion ensures each blog adheres to a user-specified word count of 200-300 words. However, neither generated blog met this limit, with the first at 566 words and the second at 1000 words, resulting in a score of 0. The Criteria Reasoning indicates that a word counter function, rather than an LLM, was used to validate this criterion.
\end{itemize}

In addition to individual scores, the Prompt Alignment Report includes a Categorical Analysis that organizes scores by instruction types, like Content, Style, Subjective, and Objective. This helps users see which types of instructions may need improvement for better alignment.

After reviewing these scores, users can make edits directly in the Prompt Panel on the left side of the screen. They can adjust any misaligned instructions and rerun the prompt to get updated scores. If users identify new constraints or realize that their needs have changed, they can go back to the \textbf{Build Your Evaluator} tab to add or adjust criteria, ensuring the evaluation process stays aligned with their evolving needs.

This process lets users improve both their prompt and criteria based on feedback from \method, creating an iterative, user-friendly approach to refining their prompts for better results.

\subsubsection{\textbf{Analyse Sample Outputs:}} In addition to generating responses, users can test alignment by uploading their own sample outputs (for example, finance blogs) through the `Analyse Sample Outputs' tab. This interface allows users to evaluate how well their examples align with the defined criteria. By testing real examples, users can confirm that the criteria effectively capture their requirements. Moreover, it can also be used to improve task understanding and requirements by updating criteria. After uploading sample responses, users are presented with a `Sample Output Alignment Report Card', similar to the `Prompt Alignment Report Card', which shows alignment scores for each uploaded example based on the saved criteria.

In the following sections, we provide the implementation details for each module.

\subsection{Implementation}\label{sec:system_backend}

\method is built on a React front end which interacts with the backend through FastAPIs. The front end is designed with close consideration given to user convenience despite the complexity of the system. The use of information hierarchy, user guides, associative colors, and visual harmony is maintained for a good user experience.
\\
In this section, we describe the implementation details of each of the four primary modules of \method. We use three types of LLMs: the \textbf{User-Specified LLM} (\( \mathrm{LLM}_{User} \)) for generating prompt responses based on user input, the \textbf{Criteria Generation LLM} (\( \mathrm{LLM}_{CGen} \)) for generating atomic instructions, criteria questions, and metadata tags, and the \textbf{Evaluator LLM} (\( \mathrm{LLM}_{Eval} \)) for evaluating responses against criteria. We use \textsc{GPT-4o} with temperature 0 for \( \mathrm{LLM}_{CGen} \) and \( \mathrm{LLM}_{Eval} \).

\noindent \textbf{\circled{1}  Evaluation Criteria Generation Module}

\begin{figure}[h!]
    \centering
\includegraphics[width=1\textwidth, keepaspectratio]{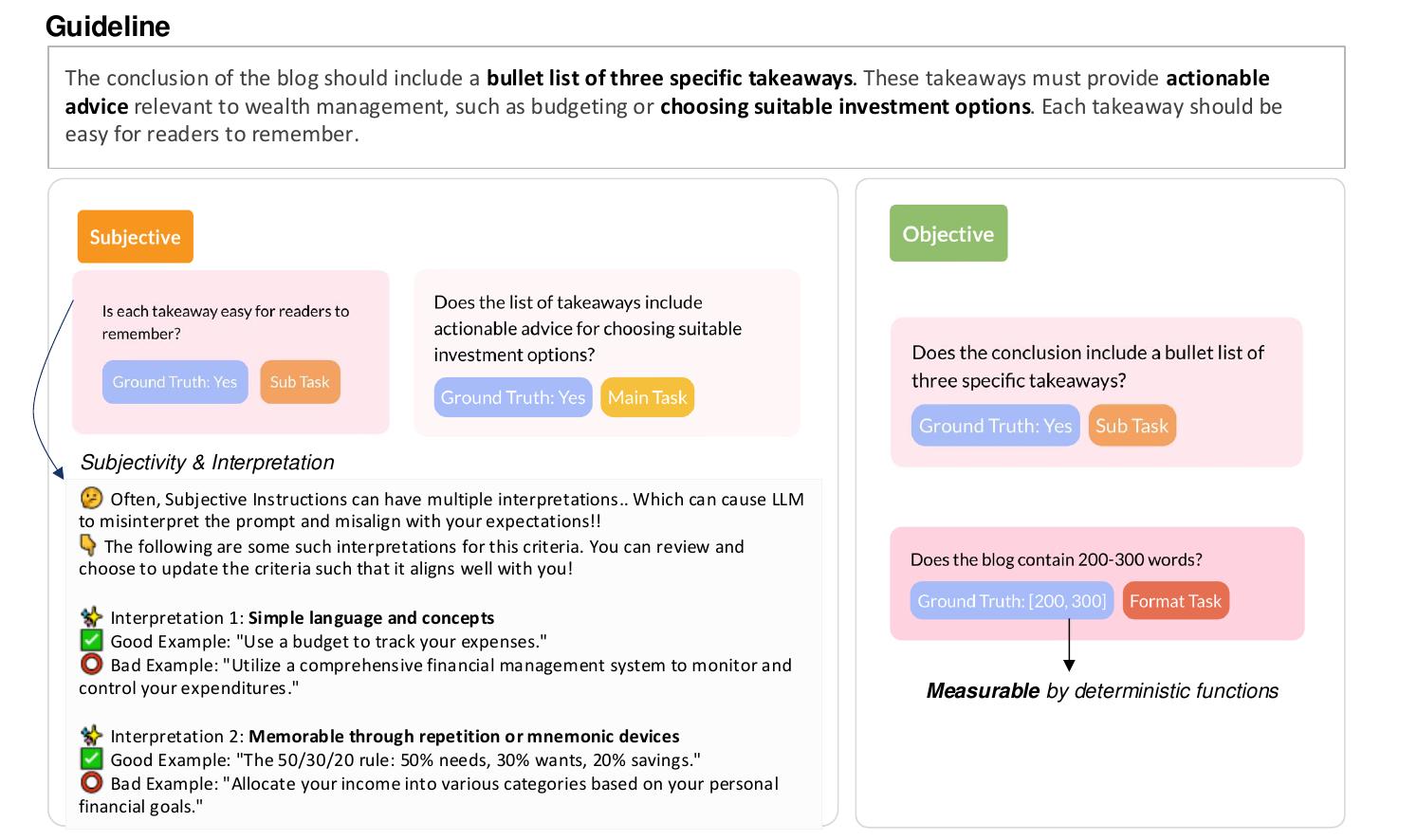}    
    \caption{This figure showcases the categorization of instructions as 'Subjective' or 'Objective' by \method. Objective tasks are clearly defined instructions, whereas subjective instructions can be interpreted in multiple ways. \method also provides the user with different interpretations of subjective instructions, along with a positive and negative example for each.    }
    \label{fig:subjectivity}
\end{figure}

In this module, the user's input guidelines are transformed into specific evaluation criteria questions. The process involves the following steps:

\paragraph{Step 1: Input Guidelines}
The user provides their prompt requirements or `guidelines'. These guidelines include instructions that the user wants to ensure are followed in the generated prompt responses.

\paragraph{Step 2:  Task Objective: } To maintain context that might be lost during decomposition, the system also extracts a task objective. This task objective is a concise summary of the main goal of the user's task. It helps contextualize each atomic instruction when generating criteria questions. For example, the task objective for the prompt in Figure \ref{fig:system_architecture} is `Generate a blog on wealth management.'  The prompt for task objective is in the Appendix \ref{sec:task_obj_gen_prompt}.

\paragraph{Step 3: Decomposing Guidelines into Atomic Instructions} 

The guidelines are decomposed into atomic instructions. Each atomic instruction represents a single, clear requirement extracted from compounded sentences that might be in the guidelines. This decomposition ensures that each criterion evaluates only a single aspect, keeping the evaluation focused, avoiding overloading criteria with multiple requirements, and reducing complexity in evaluation. The prompt for decomposing into atomic instructions is in the Appendix \ref{sec:atomic_inst_gen_prompt}.

\paragraph{Step 4:  Criteria Question Generation}


For each atomic instruction, the system rephrases it as a question,  which we refer to as a `criteria question'.  A criterion consists of:
\begin{itemize}
    \item A \textbf{\textsf{question}} that  checks if the generated prompt response meet its corresponding atomic instruction from the guidelines. 
    \item The \textbf{ground truth} expected answer to the criteria question based on the user's guidelines. These can be either `Yes'/`No', a specific number or a range (e.g. [200-300]).
    \item The inferred \textbf{priority} from the guideline. We categorize a criterion into one of the three: \begin{enumerate}
        \item `Main Task', if the criterion corresponds to an instruction from the guideline which is fundamental to the task's core logic. These also include instructions with explicit emphasis (e.g., using phrases like `make sure' or `important').
        \item `Sub Task', if the criterion corresponds to an instruction that is added to support the `Main Task' including further details or context.
        \item `Format Task',  if the criterion corresponds to an instruction that are about formatting or style of the generated responses. 
    \end{enumerate}
 Examples of these priority tags, along with criteria based on a guideline, are shown in Figure \ref{fig:subjectivity}.
\end{itemize}


 To enable automatic evaluation, we categorize each criterion based on how it should be evaluated, guiding the \( \mathrm{LLM}_{Eval} \) to select the appropriate method. Criteria are grouped into two main categories:
 
\begin{enumerate}
    \item  \textsf{Measurable Criteria} can be evaluated using  deterministic code-based functions without using \( \mathrm{LLM}_{Eval} \). Examples include word counts, sentence counts, or checking for a specific keyword. To handle more specific requirements, we add:
    \begin{itemize}
        \item  \textsf{Layered Measurable Criteria} require a two-phase evaluation process. For example, for the guideline: \textit{The conclusion of the response should be under 50 words}, \( \mathrm{LLM}_{Eval} \) first identifies the conclusion section, then applies a measurable check (word count) to verify its under 50 words. 
        
    \end{itemize}
     The ground truths for measurable criteria are either specific numbers or range of numbers. 
     
    \item \textsf{Descriptive Criteria} cannot be evaluated through simple checks as the other two categories, and are used evaluated directly by \( \mathrm{LLM}_{Eval} \). The ground truth for these are binary yes/no answer. 
    
\end{enumerate}

There are additional tags like subjectivity cues and the theme of the criteria (content/style), which further help in seeing a comprehensive alignment report. These details are provided in the Appendix \ref{sec:addtional_criteria_metadata}. Each criterion is generated using the task objective and atomic instructions from Steps 1 and 2, with prompts detailed in Appendix \ref{sec:eval_criteria_gen_prompt}. Metadata for each criterion is generated at the criterion level, with prompt details available in the Appendix \ref{sec:metadata_gen_prompt}.

\noindent \textbf{\circled{2}  Update Criteria Module}

This module allows users to delete, edit, or add criteria and accommodates evolving requirements as seen in Section \ref{sec: build_your_eval}). Once users \textit{save criteria}, these updates are used in the next evaluation, allowing for iterative refinement as users gain insights through \method.

\noindent \textbf{\circled{3}  Prompt Response Generation Module}

This module generates multiple responses based on the user’s prompt draft, using \( \mathrm{LLM}_{\text{User}} \) and the number of responses specified by the user. 

\noindent \textbf{\circled{4}  Alignment Report Card Module}

This module evaluates the generated responses against the saved criteria and presents the results in an interactive report, helping users understand how well the responses align with their requirements.


\paragraph{Determining Alignment} A criterion is considered \textbf{aligned} if the \( \mathrm{LLM}_{\text{Eval}} \)'s generated criteria answer matches the user-provided ground truth. We compute the alignment score for each response and criterion pair using the following method:
\[
\text{Score} =
\begin{cases}
1, & \text{if \( \mathrm{LLM}_{\text{Eval}} \)'s criteria-response answer matches ground truth (or within acceptable range)} \\
0, & \text{otherwise}
\end{cases}
\]

We then calculate the \emph{alignment percentage} for each criterion by averaging the scores across all generated responses:
\[
\text{Alignment\% for Criterion} = \left( \frac{\sum_{k=1}^{N} \text{Score}_{k}}{N} \right) \times 100\%
\]

where \( N \) is the total number of generated responses, and \( \text{Score}_{k} \) is the score for response \( k \). This alignment percentage is displayed next to each criterion in the UI, indicating the proportion of responses that met the criterion.

\paragraph{Aggregate Alignment Scores}  In the UI, the Alignment Report begins with an overall alignment score, shown as a pie chart with proportion of aligned versus misaligned criteria. This score is calculated as:
\[
\text{Overall Alignment Score} = \left( \frac{\text{Number of Criteria with 100\% Alignment}}{\text{Total Number of Criteria}} \right) \times 100\%
\]
where criteria with 100\% alignment are those where \emph{all} responses received a score of 1.

Next, for each instruction from the guidelines, the report (in UI) shows how many associated criteria are aligned. For example, ``1/2'' means there are two criteria linked to that instruction, and one of them is aligned. This provides a clear view of alignment at the instruction level.

Finally, at the bottom of the report, we also present alignment scores based on different categories (content or style and subjective or objective) , calculated as:
\[
\text{Alignment Score for Category} = \left( \frac{\text{Total Aligned Criteria in Category}}{\text{Total Criteria in Category}} \right) \times 100\%
\]
In this manner, we compute alignment and provide users with an interactive report for both prompt responses or their sample outputs.

\section{User Evaluation}

To evaluate how Prompt Engineers make use of \method, we conducted a user evaluation study. We wanted to find answers to the following research questions inspired by our design goals - 
\begin{itemize}

    \item \textbf{(EQ1)} How do Prompt Engineers utilize \method to improve their prompts?
    \item \textbf{(EQ2)} In what ways does \method help Prompt Engineers identify misalignments compared to traditional approaches?
    \item \textbf{(EQ3)} What are the key benefits and challenges (both usability and conceptual) of using \method for prompt improvement?

\end{itemize}

\subsection{Participant Recruitment }

To evaluate \method, we conducted a focus study with 8 participants experienced in working with complex prompts (usually over 10 instructions). This design allowed us to closely observe how prompt engineers navigate and use \method to enhance their prompts. Participants were industry practitioners recruited from an IT company via snowball sampling.\footnote{https://methods.sagepub.com/foundations/snowball-sampling}. 
Sampling criteria focused on individuals experienced in crafting long-form prompts for models from OpenAI, Meta, and others, across various use cases such as question answering, content generation, and LLM evaluation. Due to the requirement for participants to have experience with long-form prompts and spend 75 to 90 minutes in the user evaluation, our recruitment pool was limited, justifying the small sample size. We used snowball sampling to find suitable participants. Study sessions, conducted via Microsoft Teams, lasted 75 to 90 minutes to accommodate the iterative nature of prompt crafting. Sessions were recorded to capture prompt histories, interactions, and survey responses for analysis, with only the research team accessing the anonymized data. Participation was voluntary, and informed consent was obtained through consent forms.

Among the 8 participants (3 women and 5 men), 4 had engineered complex prompts for product engineering use cases, while the other 4 had industrial research applications. All participants had experience working with complex prompts to achieve desired outcomes. To maintain anonymity, we refer to participants as P[n] throughout this work.
\subsection{Study Design}

The study aimed to evaluate how prompt engineers use \method to identify misalignments and enhance their prompts. We explored the user mental model when interacting with CoPrompter, noting differences from traditional prompt improvement methods across multiple responses.

The study consisted of two parts. First, a pre-study interview gathered insights into participants' experiences with complex prompts, traditional design approaches, and alignment evaluation methods. This revealed that traditional prompt creation is time-consuming, often requiring over 10 iterations for manual response inspection. Given the impracticality of replicating this process, we relied on interviews for qualitative comparisons with \method. 

In the second part, participants were shown a demo of \method and then asked to complete a structured task. They were given a problem statement to create requirements for a feature on their product website that generates articles using LLMs for visitor engagement. This content generation use case was selected for its complexity and popularity in the formative study. The full task can be found in Appendix \ref{sec:formative_study}. Participants could choose their domain and idea, with 5-10 minutes allocated for brainstorming article guidelines. Over 40-45 minutes, participants explored \method to craft and iterate on their prompts, thinking aloud during the process. They later answered post-study questions about their experiences, discussing \method's reliability, comparison to traditional methods, workflow integration, challenges, and suggestions for improvement. Additionally, participants completed a System Usability Scale (SUS) questionnaire to rate their experience with the tool.

\subsection{Data Analysis}
The recordings from the user evaluation sessions were transcribed. The recording transcriptions were analyzed using Thematic Analysis approach \cite{ThematicAnalysis}. The transcripts were coded by the research team and the codes were collected, discussed, and categorized over three iterations using the whiteboarding tool Miro \footnote{https://miro.com/app}. SUS Scores were computed and categorized for a better understanding of user feedback. Our findings are motivated by our data analysis.

\section{User Evaluation Findings}

\subsection{\textbf{How do prompt engineers use \method to improve LLM Alignment to their Prompts?}}

As mentioned before, participants were asked to craft their own guidelines for an article generation use case on a topic they liked. The participants chose topics ranging from the need for mental well-being to the importance of investing in mutual funds. An example of a guideline crafted by a user is in Figure \ref{fig:prompt-p3}

\begin{figure}[ht!]
    \centering
   \includegraphics[width = 0.7\textwidth]{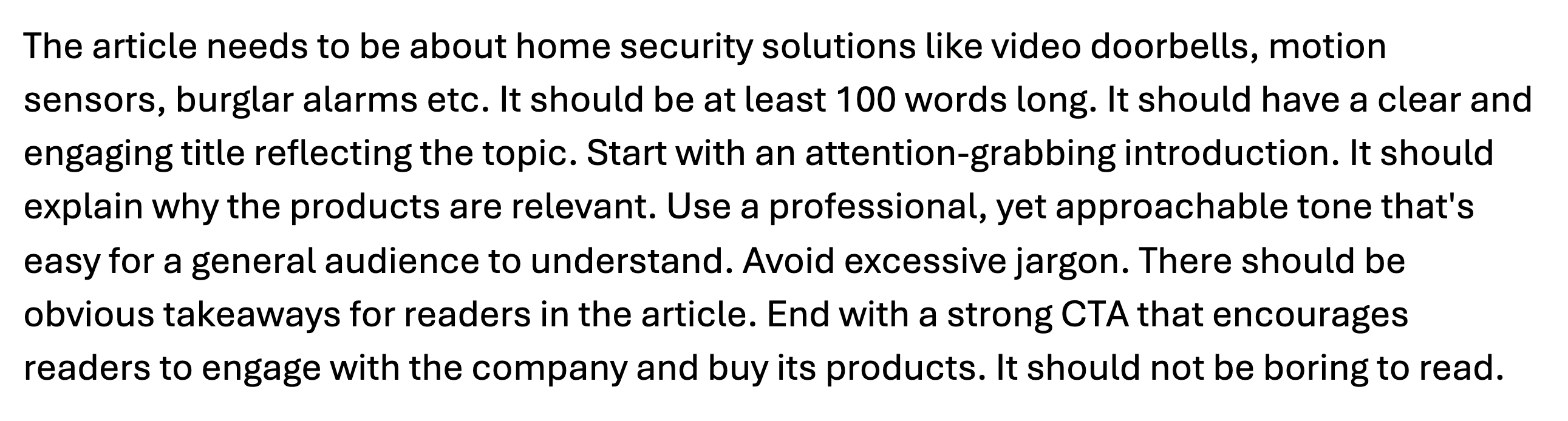} 
    \caption{Prompt Guidelines prepared by P4
    }
    \label{fig:prompt-p4}
\end{figure}


The complexity of prompt guidelines ranged from 12 to 20 instructions across participants. We observed that our participants mostly followed a consistent user flow of interacting with \method. In the following section, we answer the EQ1 and EQ3.

\subsubsection{\textbf{Building the Evaluator} using Criteria Questions:}
Once the participants had their prompt guidelines ready, they inputted them in the 'Build Your Evaluator' Tab. They were presented with a breakdown of their guidelines into criteria questions. The participants understood the concept of criteria questions by referring to the information on the interface. All the participants started their criteria review process by checking if all the conditions within their guidelines were accurately converted into criteria questions. In various instances, participants chose to make some modifications to the generated criteria list. The tool allowed participants to edit the contents or language of criteria questions, change ground truth values, delete a criteria question, and add a criteria question. We observed that in the first pass of the criteria list, the participants did not make any edits to the content or language of existing criteria questions. Participants mentioned that seeing their guidelines broken down into chunks helped them understand their requirements better and identify requirements that were missing or needed to be updated as per the \textit{ideal outcome in their minds}. For example, after reviewing the criteria list, P4 realized they wanted to revise one of their instructions, as they hadn't given it much consideration when initially drafting the guidelines. P4 stated- 
\begin{quote}
    \textit{One major problem people often face while curating prompts is that they know what they desire in the output, but they cannot break it down into different chunks, different criteria themselves to write a prompt, at least on the first go, so they might have some basic thoughts that at least these guidelines should be followed. But coming up with a bunch of criteria, the way this tool is doing that is, I think, amazing. So even the generation of criteria itself, I think is very helpful for a prompt engineer to look at these aspects and tailor the prompt accordingly.}
\end{quote}

At this stage, participants frequently added new criteria questions after identifying additional requirements. For example, P6 realized a poem should be included at the end of their response and added it using the 'Add Criteria Question' option. P3 also added checks to prevent hallucinated content. In some cases, participants found certain criteria questions unimportant or redundant and chose to delete them. For example, while testing, P5 saw two criteria with similar meanings- [C1]\textit{Does the article attract photographers and backgrounds?} [C2]
 \textit{Does the article attract photographers and backgrounds to engage with the app?}
 \\
 In this case, they deleted the first criteria to keep their criteria list optimal. Participants also updated ground truth values for criteria questions during their review, correcting inaccuracies from the Criteria Generation Module or adjusting requirements, such as changing a fixed numerical ground truth to a range of allowed word lengths. They explored other elements of the Build Your Evaluator interface and read interpretations for subjective instructions. However, P2 found the priority tags arbitrarily defined and misaligned with their priorities. After thoroughly inspecting the generated criteria, participants expressed confidence in the criteria list, saved their edits, and proceeded to analyze their prompt. Overall, participants felt the generated criteria list aligned with their requirements and appreciated the ease of editing and modifying it to better suit their needs. P5 said -
 \begin{quote}
 \textit{
     I really like the feature of modifying a criterion. Because it is particularly helpful in my case like you saw (referring to deleting criteria and adding criteria). So I think that can be like a helpful thing in general also to many people.}
 \end{quote}

\subsubsection{Analysing Prompt Response Alignment:}
\begin{figure}[ht!]
    \centering
    \vspace{-2em}
   \includegraphics[width = 0.7\textwidth]{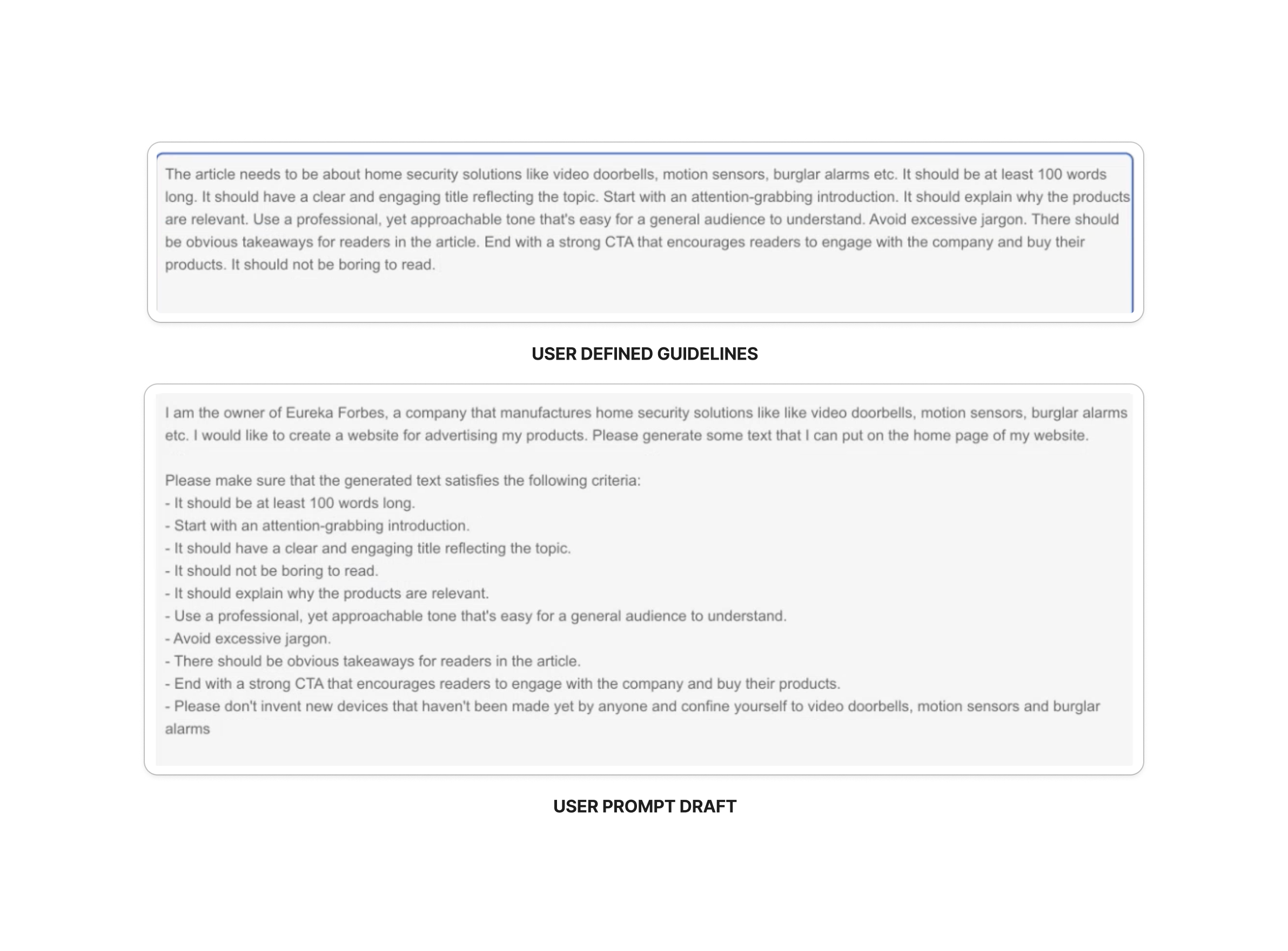} 
   \vspace{-2em}
    \caption{P3's prompt guideline for 'Build Your Evaluator' Tab and prompt draft for 'Analyse Your Prompt' Tab
    }
    
    \label{fig:prompt-p3}
\end{figure}

The participants moved to the 'Analyse Your Prompt' Tab after finalizing the criteria list. At this stage, some participants took some time to structure their prompt guidelines into a prompt-like format. For instance, P7 added additional context to the guidelines and structured the prompt. P3 broke the guidelines into bullets. Whereas, P8 did not make any changes and fed their original guidelines as prompt, as can be seen in Figure \ref{fig:prompt-p3}. Participants chose the number of responses they wanted to evaluate which varied from 5 to 10. Participants were advised to limit criteria to 10 to avoid high response generation latency due to study time constraints. Post this, participants saw the Prompt Alignment Report. The overall alignment pie chart set the context by highlighting the total number of misalignments across criteria. As the participants began to inspect the per-instruction alignment scores, they also opened responses to see the kind of responses generated by the model. It was highlighted that being able to see the responses helped the participants evaluate if \method's evaluation aligned with their preferences. Moreover, the tool's ability to provide detailed reasoning behind each score allowed participants to understand the rationale of the evaluation. This feature was highly appreciated, as it helped build trust in the evaluation process and enhanced transparency. Some participants also expressed concerns over using LLM as a judge, in this case, reasonings helped them get a transparent view of where the evaluation might be failing. As P3 stated - 

\begin{quote}
    \textit{I think the reasoning column right, I think that would be very useful even in the case of hallucinations. Sometimes the model hallucinates or does not interpret the question as you do, for that reasonings really help.}
\end{quote}

As the participants saw scores of misalignment, they edited the prompt in the left panel to modify the misaligned instruction. Participants also came up with new requirements as they saw the responses and added them in the prompt. For example. P2 observed 0\% alignment for the word length they had defined (50 to 100 words). To address this, they laid more emphasis on the instruction by adding phrases like 'Make sure' before the word length instruction.
\\
There were instances of alignment scores not aligning to how the participant would evaluate a response for a criteria, the participants were able to identify these cases by seeing the reasonings provided to justify the score. Based on this, they made a note of the changes they needed to do in the criteria questions to make them align with their requirements. Participants also discovered new checks they would want to make on their responses based on new requirements. After going through the alignment scores, participants went back to the 'Build Your Evaluator' Tab and updated the criteria questions again. 
For instance, P5 changed the language of the criteria question \textit{'Does the response engage photographers from all ages and backgrounds?'} to \textit{'Does the article mention how the app would be relevant for older photographers while also highlighting features that will attract younger photographers?'} as on inspection, they realized that the original criteria did not capture their requirement well. Participants also changed ground truths and added newer criteria. For example, P7 added a new criteria 'Does the article contain emoticons?' after seeing some responses where they expected emoticons. 
\\
Participants then re-evaluated their updated prompts against the revised criteria list, with some increasing the number of responses. P1 noted that more responses could raise the chances of misalignment, prompting further testing. Overall, participants saw improved alignment scores and confirmed that the responses better matched their requirements.

\begin{figure*}
  \includegraphics[width=\textwidth]{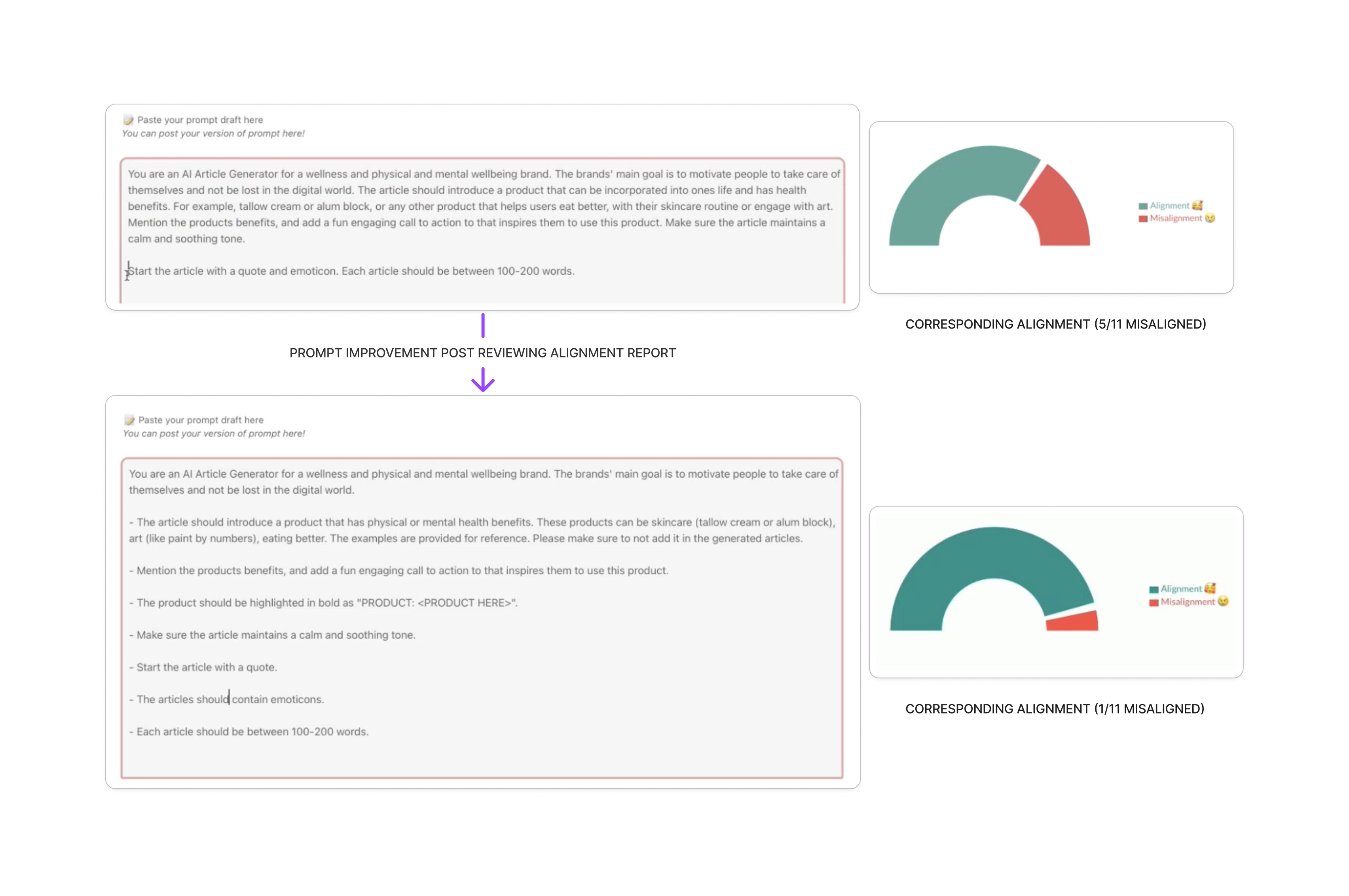}
  \caption{P7 alignment performance improved on making prompt edits (adding instructions, changing format, updating older instructions based on alignment scores) and updating the criteria list to align it better with their requirements.}
\end{figure*}

Some participants also used the 'Analyse Sample Outputs' Tab to explore \method's capabilities. P8 used ChatGPT to generate a prompt on their guidelines and fed it as a sample response to be evaluated. P3 observed a high alignment for the prompt they crafted using \method, so they used the 'Analyse Sample Outputs' Tab to explore how \method behaves for undesirable responses.


\subsection{\method as a \textit{Helping Hand} in Prompt Improvement Workflows}
 The following section answers the EQ2. Overall, participants expressed that \method can be a useful addition to their prompt improvement workflows. P3 established -

\begin{quote}

\textit{Basically, I manually check the effects of the changes that I made in the prompt, but if I have a helping hand from a tool like \method, I can check the scores, edit the prompt, and recheck the scores that it has provided, and improve where the score is low.}
\end{quote}
We observed the following broad themes on the participant user experience with \method:

\subsubsection{Criteria Generation - Enabling Control over Evaluation and Evolving Requirements}
Participants expressed that the criteria list is a great \textit{\textbf{starting point}} as a heuristic, which validates findings from literature \cite{shankar2024validates}. They valued the flexibility to modify the generated criteria throughout the evaluation process. Seeing the criteria list for the first time, or after reviewing alignment scores, often led them to identify additional requirements they hadn’t initially considered. This adaptability gave them a greater sense of control, allowing the criteria to evolve in parallel with their evolving requirements and insights. It also enabled in discovery of new requirements. As a participant reflected-

\begin{quote}
    \textit{So even if I keep aside the second part where the alignment scores are shown, the generation of criteria itself, I think is very helpful for prompt engineers as they can look at these granular aspects of their requirements and tailor the prompts accordingly. And of course it comes with the added advantage and the icing on the cake, which is a LLM also scoring your responses on these criteria. So it is like the \textbf{next level of automation}.}
\end{quote}

Overall, all the participants approved using the generated criteria list as an evaluation metric after making edits to it.

\subsubsection{Alignment Report Card – Streamlining Prompt Refinement Compared to Conventional Methods}

Participants compared \method to their traditional prompt improvement methods. Participants recalled that models usually fail at tasks related to format constraints, overlooking specific instructions, assuming some conditions on their own, misinterpreting instructions, etc. While there are many prompt engineering guides, prompt engineers usually resort to hit-and-trial methods to see if responses align with them. Additionally, making changes in one instruction could also harm the alignment of previously aligned instruction requiring constant rechecking. A participant said - 

\begin{quote}
    \textit{Many times, I don't know why I'm doing what I am doing when I am making changes in my prompts, but I just try to see if the responses get better, For example, I start trying some random things or just changing the order of instructions. like there's a lot of trial and error.}
\end{quote}

Addressing this, participants could pinpoint where exactly they needed to make a change to enhance alignment using \method. As P1 emphasized, \method shed light on \textit{what's working} and \textit{what's not working} in the prompt. While there were instances of participants not agreeing to the evaluation score, the reasonings helped them understand why \method gave that score. This helped them update the evaluation and the prompt accordingly making the process more \textit{systematic} and \textit{clean}. Overall, participants felt that \method would be a valuable addition to their workflows for prompt improvement. P2 said-

\begin{quote}
    \textit{Traditionally, It's more about trial and error, but it's not structured. So I think adding the structure to it right and having a structured way of analyzing the prompt that - actually helps out much more than just trial and error or even asking the model to optimize a prompt. So just the structure itself makes it easier to find out which part of the prompt we need to modify.}
\end{quote}

Manual inspection of multiple responses to identify misalignments was a common challenge across all participants. While some had quantitative checks for response format in their traditional pipeline, they still had to manually verify if other instructions were followed. Participants also mentioned setting up separate pipelines for quantitative checks and using code IDEs to review each instruction within complex prompts, which was cumbersome. \method offered a significant improvement by providing a one-stop solution for checking misalignments. As P3 recalled—

\begin{quote}
    \textit{I just edit my prompt somewhere in the middle of a Python file and then start a server and test it for many responses. And then dump them in some file, and then read those responses manually. And so basically I don't have one UI or tool for this, I have to do everything in the IDE basically which becomes chaotic.}
\end{quote}

Participants also appreciated not having to manually verify the alignment of every instruction across the generated responses. It commonly came up that manual inspection of misalignments takes a lot of time and is a tedious process. \method breaks the prompt into criteria and conducts an evaluation across all the responses. This came up as an aid to save time and streamline the evaluation process. A participant reflected - 

\begin{quote}
    \textit{It's like usually hard to manually inspect like, say 15 criteria on say 20 or 30 responses right? So that's a big problem that's solved, it will save a lot of time.}
\end{quote}

Participants also appreciated the User Interface of the tool. The interface seemed intuitive and easy to follow. Participants stated that the user flow of the interface is similar to how they usually go about improving prompts. This included using guidelines to make a prompt and generating multiple responses which are then evaluated manually. Participants appreciated the tags and color coding to better understand their instructions.

\begin{quote}
    \textit{A lot of the things in this tool are very easy to follow. I mean, initially when a new user comes to this interface, obviously there are going to be some questions, but if you consider a user who is going to be regularly using this. I think things are really, really easy to follow and adapt for one's own use cases. So I think that is really commendable. I notice the use of specific colours to indicate different things and all that actually stands out in the subconscious mind, and this is something that definitely is worth appreciating.}
\end{quote}

Lastly, we observed that participants could also identify some nontrivial applications of \method inspired by their use cases. P3 mentioned that \method could be used to assist human annotations by reducing the effort of evaluation through human annotation across all responses to just needing human input at the criteria review stage. P1 also pointed that \method can also serve as a tool to support the interpretability of the model response behavior as it highlights what types of instructions a model is not able to consistently follow. Interestingly, various participants expressed interest in expanding the \method architecture to text-image models to get alignment scores by evaluating images. P7 stated that \method could also be used as a way of documenting prompt evolution journeys for monitoring and presentation purposes. They said that '\method scores' can be used as standards to pass prompt qualities for a use case. Overall, participants were intrigued by the concept of quantifying alignment and found it relatable to challenges they frequently encounter in their daily use cases.

\subsection{Painpoints and Suggestions }

\subsubsection{Challenges in navigating \method}

The following section answers EQ3. Participants navigated \method with ease but expressed initial hesitation about using LLMs to evaluate certain questions. Coming from a prompt engineering background, opinions on LLMs as evaluators varied, with some initially expressing distrust. However, features such as viewing generated responses, understanding the reasoning behind alignment scores, and analyzing sample responses helped alleviate their doubts and build trust in the evaluation process.
\\

P1 and P2 felt that \method would be challenging for more objective tasks like code evaluation, suggesting a version that breaks down evaluation into chunks, similar to criteria lists. Participants sometimes struggled to understand the reasoning behind the tags associated with each criteria question. For instance, P4 noted that the instruction "it should sound exciting" could affect both content and style, while P8 was confused about the distinction between Objective and Subjective tags. This indicates a need for clearer explanations of these concepts. P1 also found the system complex, requiring the learning of various concepts, though other participants did not face similar challenges. Addressing system complexity through improved explainability and providing learning guides was suggested by P1.
\\
P8 observed that when creating a prompt, the significance of each instruction can vary considerably. They pointed out that an overall alignment score of 9/11 might not capture their judgment on alignment, as misalignments in less critical instructions would matter less to them compared to misalignments in more important ones. P8 suggested incorporating the relative weight of each instruction within the prompt to produce a more accurate overall alignment score that better reflects the prompt engineer's priorities.
\\
Additionally, participants experienced wait times of up to a minute during evaluations with a high volume of responses, primarily due to API call latency. P1 referred to this wait as the "cost" of the ease \method provided in response evaluation. Despite the latency, participants remained patient and did not view it as a significant challenge.
\\
All participants except P8 stated they could integrate \method's current flow into their existing prompt improvement pipelines. In contrast, P8 preferred receiving alignment reports through a command-line interface, as they typically run API calls via a terminal, which would reduce onboarding friction. They expressed interest in an SDK version of \method.

\subsubsection{Suggestions and Points of Improvement}

On using \method, various participants suggested adding prompt improvement suggestions using LLMs and \method scores. Participants stated that prompt improvement suggestions will further make their process easier as they will also have some actionable suggestions for improvement corresponding to alignment scores. P4 suggested guideline suggestions by detecting the intent of the task. P4 stated that users sometimes forget to include some obvious guidelines or do not think enough, implicit guidelines suggestions could assist in such cases. P1 suggested having prompt templates available to help create effective prompts. Overall, participants mostly seemed interested in also knowing actionable ways to improve the prompt after identifying the area of improvement.
P3 suggested that using reasoning in prompt alignment reports can also help detect hallucinations if the reasoning is grounded in some documents provided by the user. 
Lastly, P8 suggested that if misalignments can be highlighted while the user is writing a prompt, like grammar errors are highlighted in Grammarly\footnote{https://app.grammarly.com/}, it will save the time spent on reviewing the prompt alignment report.

\subsection{System Usability Scale (SUS) Survey}


\begin{figure}[h]
    \vspace{-20mm}
    \centering
    \includegraphics[width=1\textwidth]{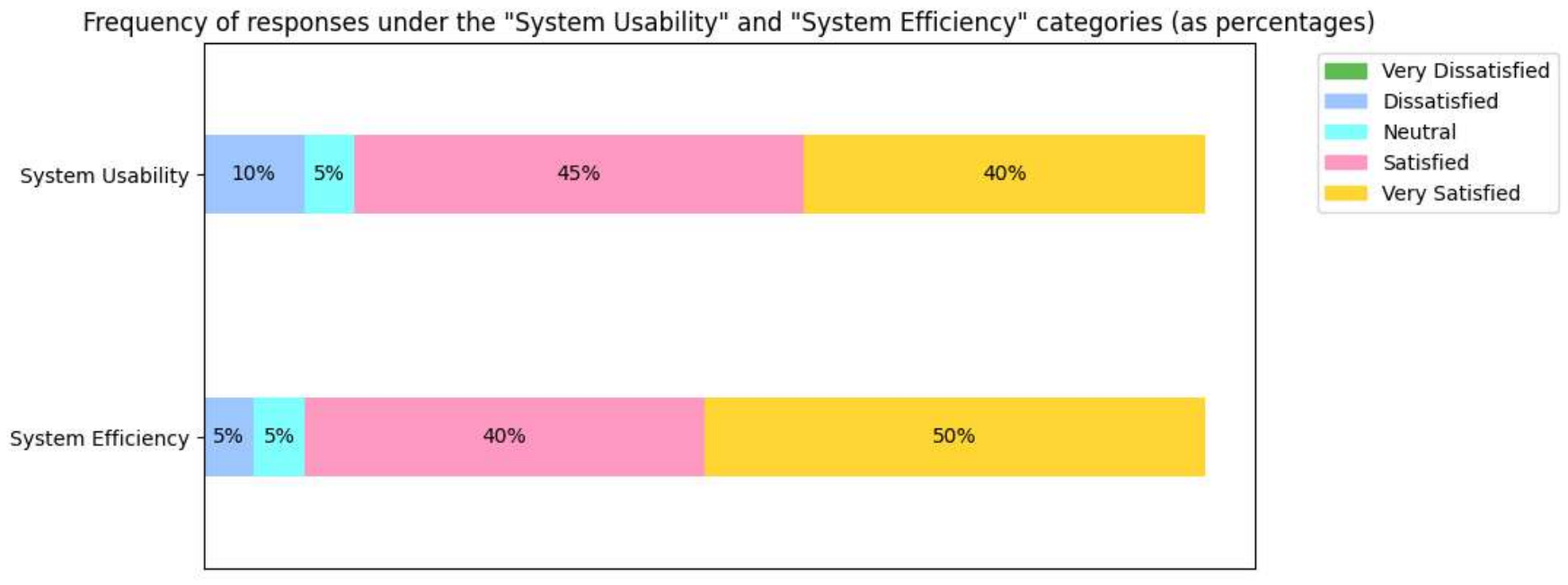} 
    \vspace{-30mm}
    \caption{Participant ratings on System Usability and Efficiency}
    \label{fig:SUS_Categories_figure}
\end{figure}

Participants completed a SUS Survey after interacting with \method, consisting of 10 questions. These were divided into two categories: System Usability (user-friendliness) and System Efficiency (goal achievement), with 5 questions in each. The full list of questions is provided in the appendix.
We aggregated the ratings for "System Usability" and "System Efficiency" categories. As shown in Figure \ref{fig:SUS_Categories_figure}, 45\% of users were "satisfied" and 40\% "very satisfied" with System Usability, reflecting strong positive feedback and frequent use of \method. For System Efficiency, 50\% rated it "very satisfied" and 40\% "satisfied," indicating the tool was easy to use, quick to learn, and provided a smooth user experience.

\begin{figure}[h!]
    \vspace{-35mm}
    \centering
    \includegraphics[width=1\textwidth]{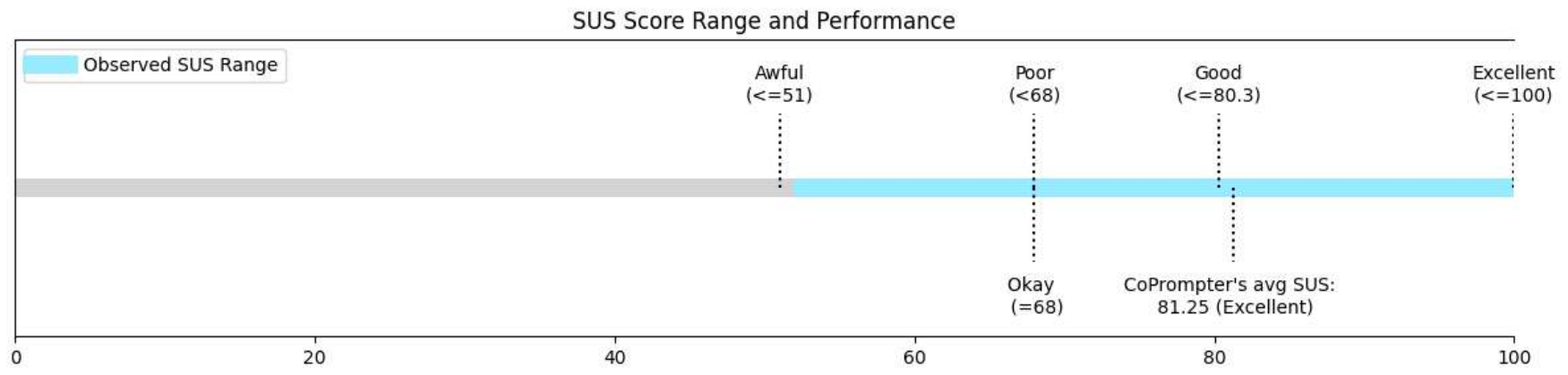} 
    \vspace{-40mm}
    \caption{SUS Score Range and Aggregate across participants}
    \label{fig:SUS_Scale_figure}
    \vspace{-5mm}
\end{figure}

Participants' System Usability Scale (SUS) scores yielded a mean overall SUS score of 81.25, positioning the usability of our tool within the "A" range of excellence, as per Sauro and Lewis's Curved Grading Scale (CGS) \cite{Sauro_Lewis_2012} as shown in figure \ref{fig:SUS_Scale_figure}. This score significantly exceeds the industry average SUS score of 68 ("Okay"). \cite{Sauro_2011}, thereby indicating that the usability of our tool surpasses that of typical software interfaces \cite{Bangor_Aaron_Kortum}. P1 expressed dissatisfaction toward system complexity which can be seen in \ref{fig:SUS_Categories_figure} as the blue bar. Our overall score suggests that users perceive our tool as both efficient and user-friendly, enabling quick task completion with minimal frustration, thereby validating the effectiveness of our design. 


\section{Discussion}

\subsection{\method as an assistant for Prompt Improvement}

Our user evaluation revealed how participants utilized \method to improve their prompts. Our observations suggest that \method can be leveraged effectively to assist prompt engineers in evaluating instruction following misalignment systematically \sethlcolor{yellow}
\hl{\textbf{(DG1)}}. Participants emphasized how \method is an improvement on their existing evaluation methods of manually inspecting multiple responses to identify misalignments which are rather time-consuming and cumbersome. Participants identified \method as a one-stop platform to evaluate multiple responses with ease, as it conducts checks over all the user requirements and provides a detailed report on instruction level alignment of responses. More so, \method enables a \textit{user in loop} evaluation such that the alignment evaluation aligns with the user's definition of alignment. The user evaluation also pointed out that prompt engineers can make systemic changes to refine their prompt \sethlcolor{yellow}
\hl{\textbf{(DG2)}} as \method pinpoints which exact instructions are misaligned, guiding users to improve prompts in a systemic manner. This also emerged as a significant improvement over traditional prompt improvement methods, where participants often relied on trial and error, unsure of where exactly to make adjustments. 
Lastly, \method supports the evolving nature of user evaluation requirements \sethlcolor{yellow}
\hl{\textbf{(DG3)}} by providing control over the Evaluation Criteria List. As demonstrated in the user studies, participants discovered new prompt requirements throughout the process of alignment review. Participants appreciated the flexibility to modify the criteria questions as per their evolving requirements.

While \method archives the goal of streamlining misalignment identification and prompt improvement, participants also identified additional potential applications. They noted that \method could help clarify their requirements by breaking down guidelines into granular instructions, aiding in drafting prompts that accurately capture intent. Furthermore, it facilitates model interpretability by providing a categorical analysis of instructions the model struggles with. Participants also mentioned its potential for tracking prompt and alignment score evolution for monitoring and presentation. Additionally, \method could act as a human annotator by conducting human-aligned evaluations of responses. Overall, \method was validated as a valuable tool for prompt engineers, with opportunities for expanding its use cases.

\subsection{Considerations for designing for Human LLM Alignment Workflows}
\subsubsection{Alignment is Dynamic}
The meaning of Human LLM alignment has evolved over time from being seen as a uni-directional concept (driven by either AI or Humans) to a \textit{bi-directional framework} where Human alignment to AI and AI alignment to Human are seen as interconnected processes \cite{TowardsBidirectionalHumanAIAlignment}. Terry et al. shed light on how the process of alignment can be broken down into - specification, process and evaluation alignment \cite{InteractiveAIAlignment}. These concepts become more complex because alignment is not a static phenomenon. Its dynamism is \textit{twofold} —varying across users and continuously evolving over time. As seen in our formative study and user evaluation, individuals have unique definitions of what passes as 'response alignment' to them. Therefore, a \textit{user-centric approach} to building systems supporting Human AI Alignment becomes critical. \method does that by enabling user control over the evaluation process of alignment. This control provides users with the agency of \textit{defining alignment} for their use case and does not assume what alignment might mean to the user. 
\\
Moreover, the meaning of alignment for a user can also evolve over time. As seen in the user studies, users came across newer requirements and expectations as their prompt evolved. Human-AI Alignment systems need to take this \textit{evolution of alignment} throughout the journey of prompt engineering into account. \method does that by enabling users to modify the evaluation pipeline at any point. We recommend future designers and researchers working in this space to be sensitive to the twofold dynamism of alignment while building systems to facilitate prompt designers in navigating prompt improvement. This could be achieved by leveraging user-driven alignment approaches that incorporate human aspirations, values, and assets in evaluation design. These human factors can be discovered by using human-centric research methods and frameworks such as participatory design, value-sensitive design, asset-based design, etc. \cite{RAMOSMONTANEZ2023101540, SurveyofValueSensitiveDesign, MethodologyOfParticipatoryDesign}.

\subsubsection{Trust as a Key Driver in Alignment Evaluation Workflows}

Human trust towards LLMs is a common factor that defines Human AI Interaction \cite{CanWeTrustAI, TrustInHumanAI, 1_Introduction, CHI2021106700}. We observed instances of mistrust and hesitation in users when they were told about the role of LLMs in the alignment evaluation. LLMs as judges can be biased, lack nuanced understanding and may misinterpret context or intent, leading to unreliable or unfair outcomes \cite{HumanCenteredDesign}. However, as shown in the user evaluation, \method was able to address this possibility of mistrust towards the alignment evaluation using LLMs through various measures. Participants were able to see the reasoning behind a particular alignment score given by the evaluator model to understand the rationale of the score. This also helped them identify instances where the score did not align with them and decided to trust it accordingly. More so, participants could also test \method on custom responses provided by them to see how alignment scores compare to their judgment of alignment for a sample response. These measures helped participants overcome doubts and mistrust towards \method's ability to identify misalignment. It also helped in building a transparent interaction with the tool. Participants also mentioned that this assisted in better interpreting the behavior of the LLMs. We highlight user trust as an important consideration while building Alignment Evaluation Workflows in order to assist users in ways that they can rely on. Since Alignment is user-defined, it becomes critical to make sure that the user is able to follow the rationale behind the alignment judgment.

\subsection{Future Work and Limitations}

The user evaluation demonstrated that \method meets the study's design goals while revealing opportunities to expand its support for prompt engineers. Future enhancements could include suggestions for prompt improvements, guideline development, and usability features like prompt templates to simplify design. We also identified the need for a nuanced approach to alignment, considering priority, weights, and implicit user intents. These areas represent potential expansions for \method and opportunities for the IUI community to further explore Human-AI Alignment. Additionally, user studies highlighted various user experience improvements for future versions of \method.
\\
This work also poses certain limitations. The evaluation of \method happened in a focus study format. We believe we can learn more about usage patterns and pain points through in-the-wild studies by exploring the different use cases users use \method for. More so, with the limitation of the sample size, we could text \method with 8 participants. We plan on expanding this as well to test it with a higher number of users with varying use cases to unravel possible applications of \method. More so, the latency in response generation required users to wait before they could the evaluation. While this did not hamper the user experience in our studies, this can be improved for a more efficient workflow.
\section{Conclusion}

In conclusion, user-in-loop methods for evaluating LLM responses can play a crucial role in improving Human-AI alignment. \method does this by efficiently identifying misalignments between LLM outputs and user-defined criteria, enabling users to iteratively refine their prompts.

The positive feedback from industry prompt engineers, reflected in high System Usability Scale (SUS) scores, underscores the effectiveness of \method in refining prompts and adapting to evolving requirements. Our contributions provide a systematic approach to pinpoint areas for improvement and offer insights into the challenges faced by prompt engineers. This work demonstrates that with the right tools and methodologies, it is possible to significantly enhance the reliability and usability of LLMs, fostering greater confidence in their application across diverse domains.

\bibliography{coprompter}

\newpage

\section{Appendix}

\subsection{Additional Criteria Metadata}\label{sec:addtional_criteria_metadata}

\begin{enumerate}
    \item \textbf{\textsf{Subjectivity Tag}}:  This attribute represents whether the atomic instruction is Subjective or Objective. Subjective terms are defined as words or phrases that can be interpreted in multiple ways. We added this attribute to highlight to the users that a particular atomic instruction might need special attention due to the present subjectivity.  If a particular instruction is tagged as subjective, we also add different interpretations of the instruction along with a positive example and a negative example of the particular interpretation.  Based on the interpretations and examples, the users can acquire a clear understanding of how to specify their requirements in the prompt in a more objective manner.

    In Figure \ref{fig:subjectivity}, the instruction, 'The conclusion of the blog should include a bullet list of three specific takeaways.' is labeled as objective, as it provides a clear requirement without any ambiguity. The instruction specifies a format (a bullet list), a fixed number (three takeaways), and a section (the conclusion).
    Meanwhile, the instruction that each takeaway should be "easy for the user to remember " is tagged as subjective. The phrase "easy to remember" can vary widely in interpretation, leading to different approaches in the generated responses. The first interpretation suggests simplicity in language and concepts, offering clear, straightforward advice like using a budget to track expenses. The second interpretation focuses on memorability through repetition or mnemonic devices, such as the 50/30/20 rule for budgeting. By providing examples for both interpretations, the user can see how the instruction might be understood in different ways, potentially causing misalignment in response generation. This helps the user refine the instruction to better match their intent.

    To avoid very close interpretations, we also generate a similarity score for interpretations on a scale from 1 to 5.  1 indicates that the interpretations are very different, and 5 indicates that they are very similar. This way, the generated interpretations can be filtered if needed.

    \item \textbf{\textsf{Question Theme}} This attribute explains the theme of the criteria question. The two different tags for the attribute are \emph{Content}, \emph{Style}. Criteria questions that check the qualitative style or overall format of the response will be of the \emph{Style} theme, for example, tone, vibe, or output format. Criteria questions that check the content of the response, that is, if certain content information is present or if the content task is followed well, will tagged as \emph{Content}.

\end{enumerate}

\subsection{Formative Study Questionnaire}
\label{sec:formative-study-questionaire}

The formative study collected insights from participants regarding their experiences, challenges while prompt engineering. Responses reflect the number of users who chose each option, providing a quantitative view of common practices and issues encountered in various application contexts.

\textbf{(Q) In what context have you previously developed tools/applications using finetuned LLMs? (Select all that apply)} 
\begin{itemize}
    \item Customer Facing Products: 17
    \item Company Hackathons: 14
    \item Personal Applications: 10
    \item Internal Products: 6
    \item Non-customer facing products: 1
    \item Company Workshops: 1
    \item POC development: 1
\end{itemize}

\textbf{(Q) What kind of tasks have you used LLMs for in the above-mentioned tool? (Select all that apply)} 
\begin{itemize}
    \item Knowledge Extraction/ Information Retrieval: 20
    \item Content Generation: 19
    \item Conversational Support through chatbots: 15
    \item Summarization: 15
    \item Personalisation: 3
    \item API Execution to perform tasks in the application: 1
    \item Instruct-guided editing workflows: 1
    \item Assistant in using the product - executing tasks for the user: 1
    \item Make charts and graphs: 1
    \item Question Answering: 1
    \item Text Classification: 1
\end{itemize}

\textbf{(Q) How do you often structure your instructions in your prompt to the LLM for your applications? (Select all that apply)}
\begin{itemize}
    \item Instructions supported with Examples: 21
    \item List of to-the-point and concise Instructions: 20
    \item List of detailed and verbose Instructions: 16
    \item Instructions using content from multiple LLM calls: 11
    \item Step-by-step demonstrations of how to generate response: 10
    \item Unstructured Instructions (paragraph format): 10
    \item Pseudocode instructions using SudoLang: 1
    \item Enriched RAG with KG query. Semantic search filtering: 1
    \item Optimizing prompts prior to use in a production tool: 1
\end{itemize}

\textbf{(Q) What are the different types of instructions you usually give in a prompt? (Select all that apply)}
\begin{itemize}
    \item Instructions about the objective task to be implemented: 26
    \item Instructions about the desired format of response: 25
    \item Instructions about the context of the conversation: 25
    \item Instructions about the desired style of response: 14
    \item Interface-oriented, constraint-based programming to express complex behaviors: 1
\end{itemize}

\textbf{(Q) What type of misalignments do you observe in the response generated by the LLM? (Select all that apply)}
\begin{itemize}
    \item Some instruction overlooked/ignored: 25
    \item Hallucinations: 21
    \item Misinterpretation of instructions: 18
    \item Assumptions/elements not instructed: 14
    \item Inconsistent instruction following: 14
    \item Overfitting to examples: 12
    \item Incomplete response: 7
    \item Misinterpretation of Time Zones: 1
    \item Irrelevant response: 1
    \item Categorization issues and failure to follow rules: 1
    \item Failure in tool usage and format consistency: 1
\end{itemize}

\textbf{(Q) Which types of instructions do you observe misalignment issues in? (Select all that apply)}
\begin{itemize}
    \item Complex, elaborate instructions: 19
    \item Format Instructions: 15
    \item Objective Instructions for conditional actions: 15
    \item Objective Instructions for content generation: 14
    \item Subjective Instructions: 7
    \item Style Instructions: 6
    \item Instructions with Examples: 5
    \item Manipulation of discrete data: 1
    \item API-related chat questions: 1
    \item Centered text when using examples: 1
    \item Missing or unclear instructions: 1
\end{itemize}

\textbf{(Q) When you encounter misalignment, what strategies do you use to get a better response? (Select all that apply)}
\begin{itemize}
    \item Identifying misaligned instructions: 21
    \item Providing more context: 19
    \item Testing over multiple responses: 19
    \item Hit-and-trial improvements: 18
    \item Adding new instructions: 17
    \item Breaking tasks into steps: 14
    \item Reordering instructions: 14
    \item Adding more examples: 10
    \item Voyager technique: 1
    \item Adding few examples cautiously: 1
    \item Small datasets or vector database issues: 1
    \item Reinforcing attention via Local COT: 1
    \item Identifying ambiguities: 1
\end{itemize}

\textbf{(Q) From the above context, which of the following modifications often improve response alignment? (Select all that apply)}
\begin{itemize}
    \item Providing more context: 16
    \item Adding new instructions: 15
    \item Breaking tasks into steps: 14
    \item Hit-and-trial language improvements: 10
    \item Reordering instructions: 10
    \item Testing multiple responses: 10
    \item Simplifying complex prompts: 1
    \item Adding rare examples: 1
    \item Upgrading to a newer model: 1
    \item Inconsistently effective: 1
    \item Restarting project: 1
    \item Example input-output tests: 1
    \item Identifying ambiguities: 1
\end{itemize}

\textbf{(Q) What challenges do you face when trying to improve response alignment? (Select all that apply)}
\begin{itemize}
    \item Prompt refinement takes time: 20
    \item Inconsistent instruction-following: 17
    \item Unexpected response behavior: 17
    \item Lack of modification clarity: 7
    \item Choosing suitable LLM model: 4
    \item Difficulty expressing intent: 4
    \item Limited LLM capacity or attention heads: 1
    \item Misalignment trade-offs in output: 1
    \item Category issues: 1
    \item Token and retrieval limitations: 1
\end{itemize}

\textbf{(Q) Which LLM do you use often for these use cases?} 
\begin{itemize}
    \item OpenAI Models: 27
    \item Other Large Language Models: 6
    \item LLama Models: 5
    \item Custom fine-tuned Models: 4
    \item GPT-4, Claude 3 Opus, Llama 3 70b, Mixtral 8x7b: 1
    \item Exploring Claude 3, Mistral: 1
    \item Mistral/Mixtral: 1
    \item Vision: 1
    \item Sonnet (personal projects): 1
\end{itemize}

\textbf{(Q) How many instructions do you usually use in your prompts?} 
\begin{itemize}
    \item More than 10 instructions: 10
    \item 5-10 instructions: 8
    \item 2-5 instructions: 5
    \item 1 instruction: 4
    \item Varies (3 to more than 10): 1
\end{itemize}

\textbf{(Q) How many attempts does it usually take to design a prompt for the desired response?}
\begin{itemize}
    \item More than 10 attempts: 14
    \item 4-6 attempts: 6
    \item 2-3 attempts: 4
    \item 7-10 attempts: 3
    \item 1 attempt: 1
\end{itemize}

\textbf{(Q) On a scale of 1-5, how aligned is the LLM response to your instructions on the first attempt?}
\begin{itemize}
    \item Rating 3 - 12 
    \item Rating 4 - 11
    \item Rating 5 - 12
\end{itemize}

\textbf{(Q) How many generated responses do you test your prompt on?} 
\begin{itemize}
    \item More than 40 responses: 14
    \item 20-40 responses: 5
    \item Less than 20 responses: 4
\end{itemize}

\textbf{(Q) What are the inputs you use while testing/evaluating your prompt?} 
\begin{itemize}
    \item Manually created dataset: 8
    \item Sample inputs + dataset: 5
    \item Sample inputs only: 3
    \item RITEWay test suite: 1
    \item Various dataset combinations: 1
\end{itemize}

\textbf{(Q) Which stages do you check for alignment to prompts?} 
\begin{itemize}
    \item Prototype: 11
    \item Prototype + Production: 6
    \item All Stages (Prototype, Production, Post-deployment): 1
\end{itemize}

\textbf{(Q) Please tell us more about your instruction styles and instruction categories used in your prompts here.} 
Various responses, including single-shot, few-shot, multi-agent architectures, constraint-based programming, specific document grounding, iterative methods, example-based formats, and custom tagging for specific cases.

\subsection{User Interface}\label{sec:UI_Pics}

\vspace{-2mm}
\begin{figure}[H]
    \centering
    \includegraphics[width=0.8\linewidth, height=1.1\linewidth, keepaspectratio]{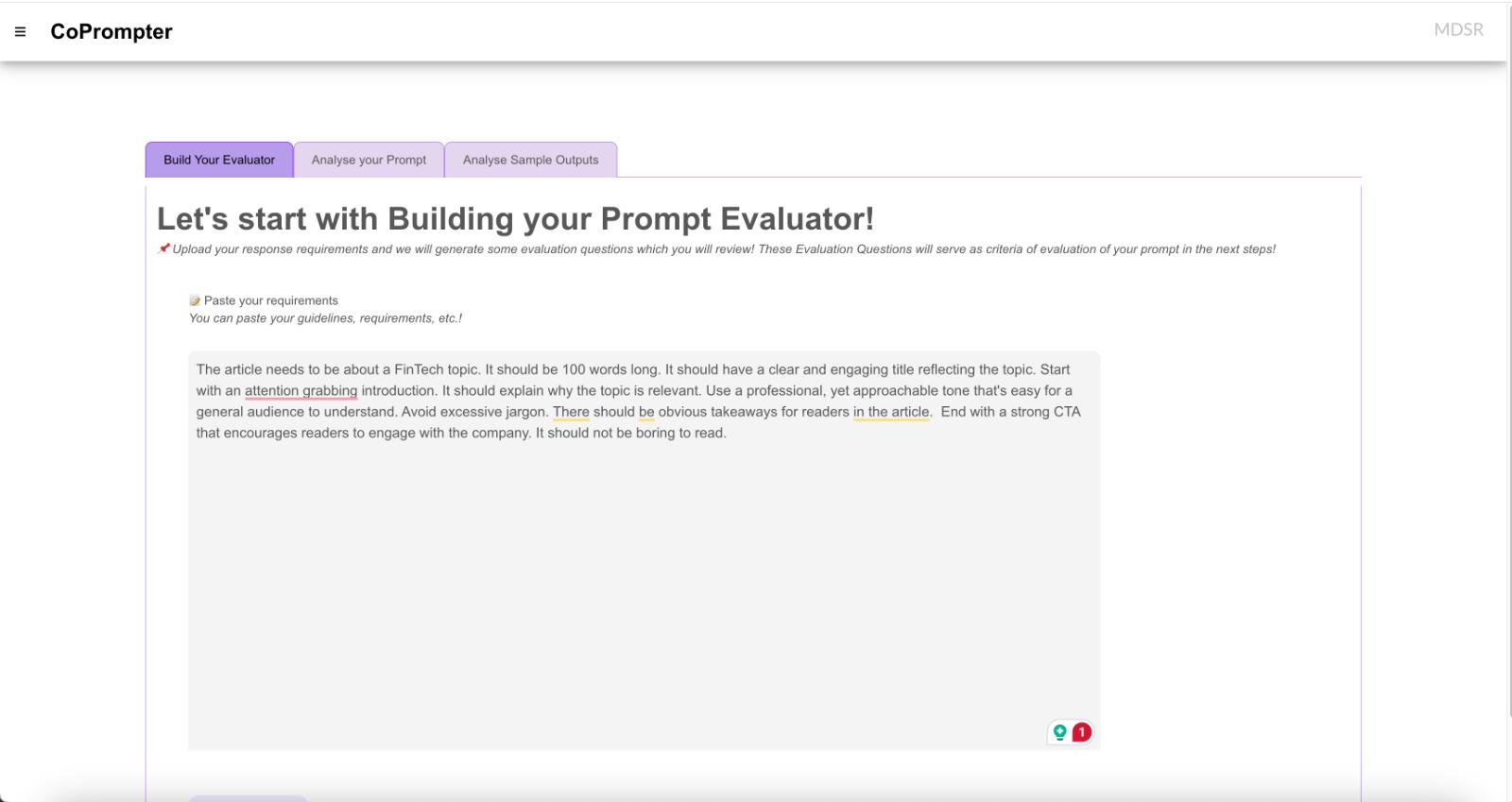}
    \caption{"Build your evaluator" tab}
\end{figure}

\vspace{-55mm}
\begin{figure}[H]
    \centering
    \includegraphics[width=0.8\linewidth, height=1.1\linewidth, keepaspectratio]{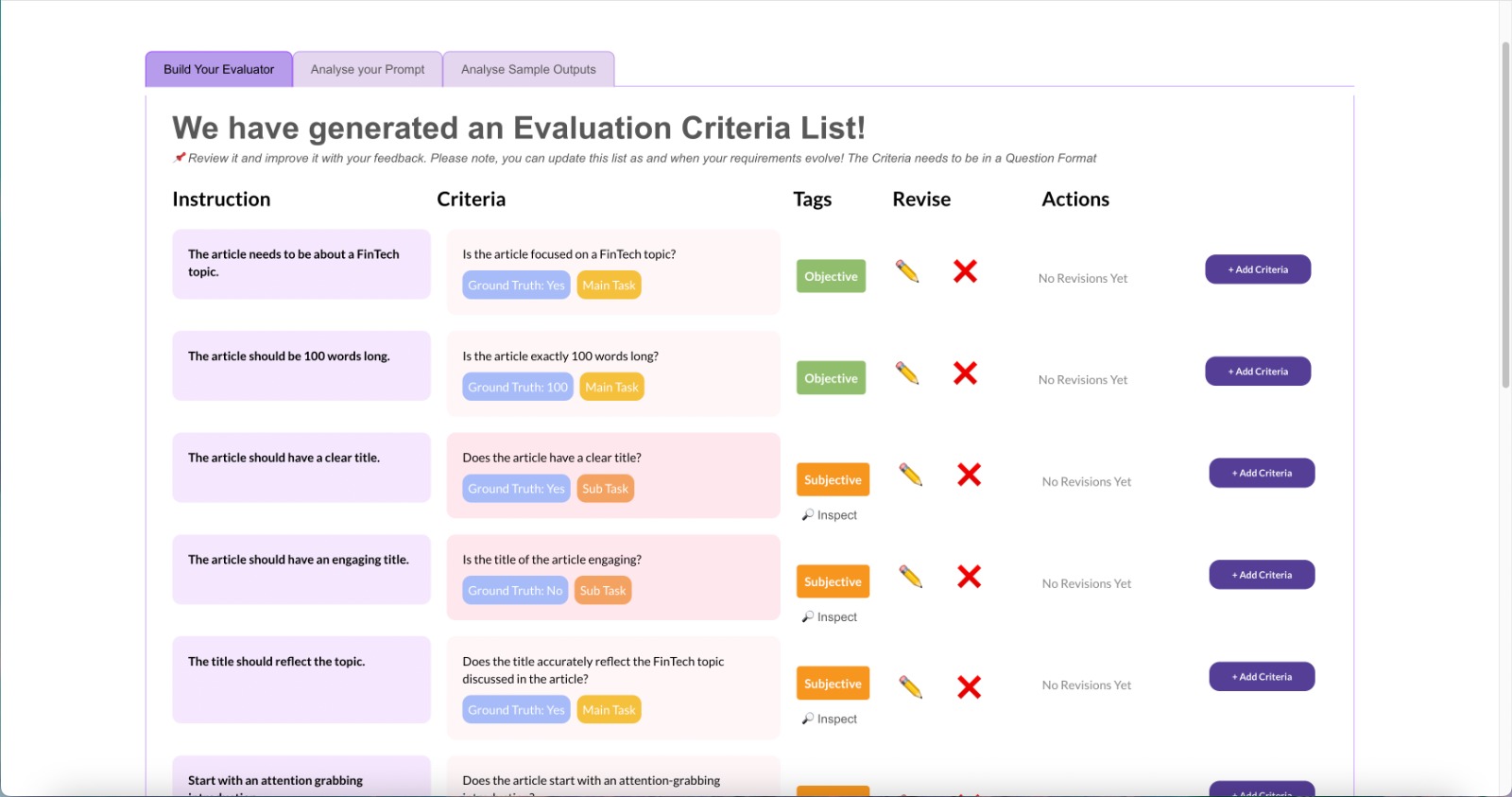}
    \caption{View criteria list}
\end{figure}

\vspace{-45mm}
\begin{figure}[H]
    \centering
    \includegraphics[width=0.8\linewidth, height=1.1\linewidth, keepaspectratio]{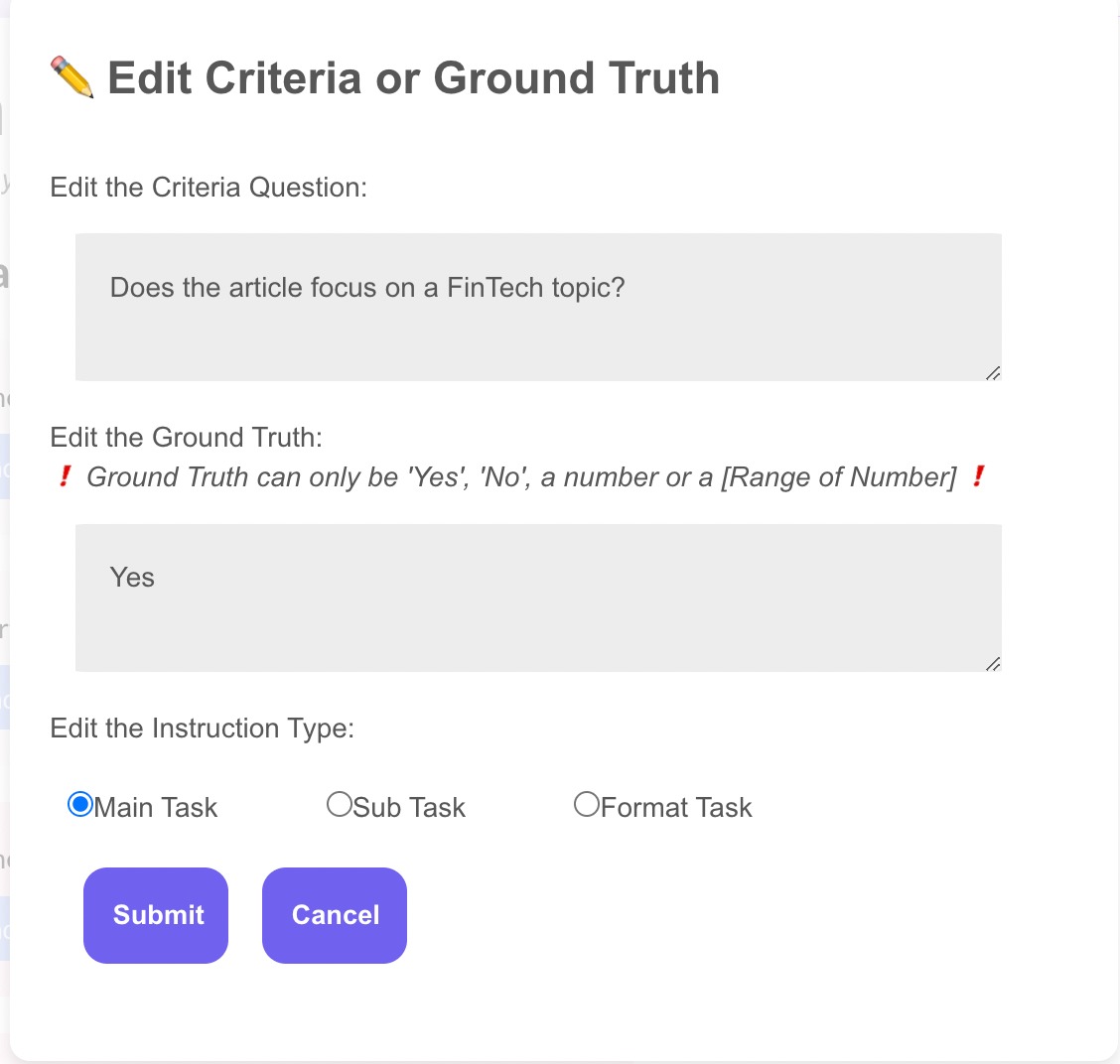}
    \caption{Edit criteria}
\end{figure}

\begin{figure}[H]
    \centering
    \includegraphics[width=0.8\linewidth, height=1.1\linewidth, keepaspectratio]{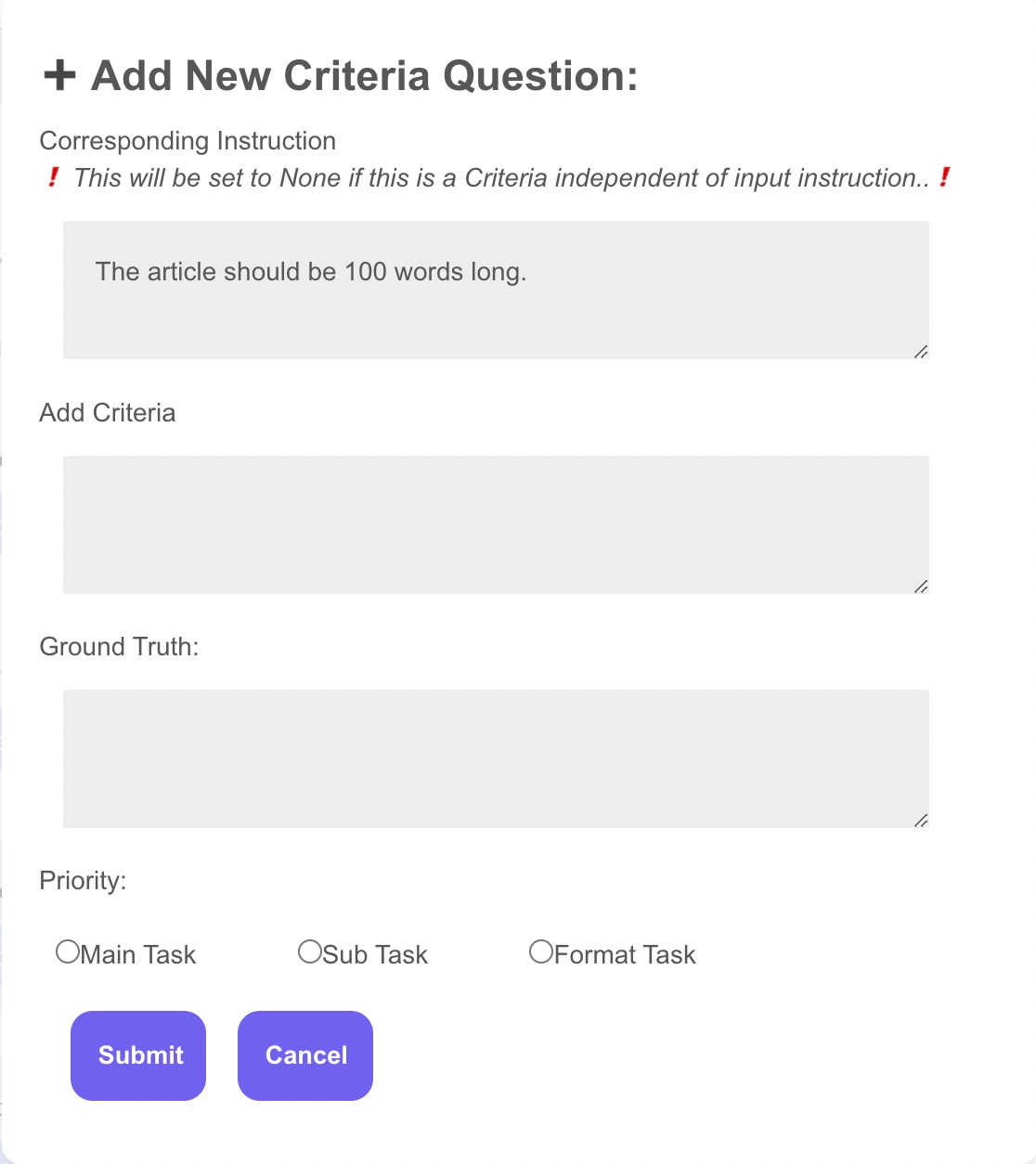}
    \caption{Add criteria}
\end{figure}

\vspace{-20mm}
\begin{figure}[H]
    \centering
    \includegraphics[width=0.8\linewidth, height=1.1\linewidth, keepaspectratio]{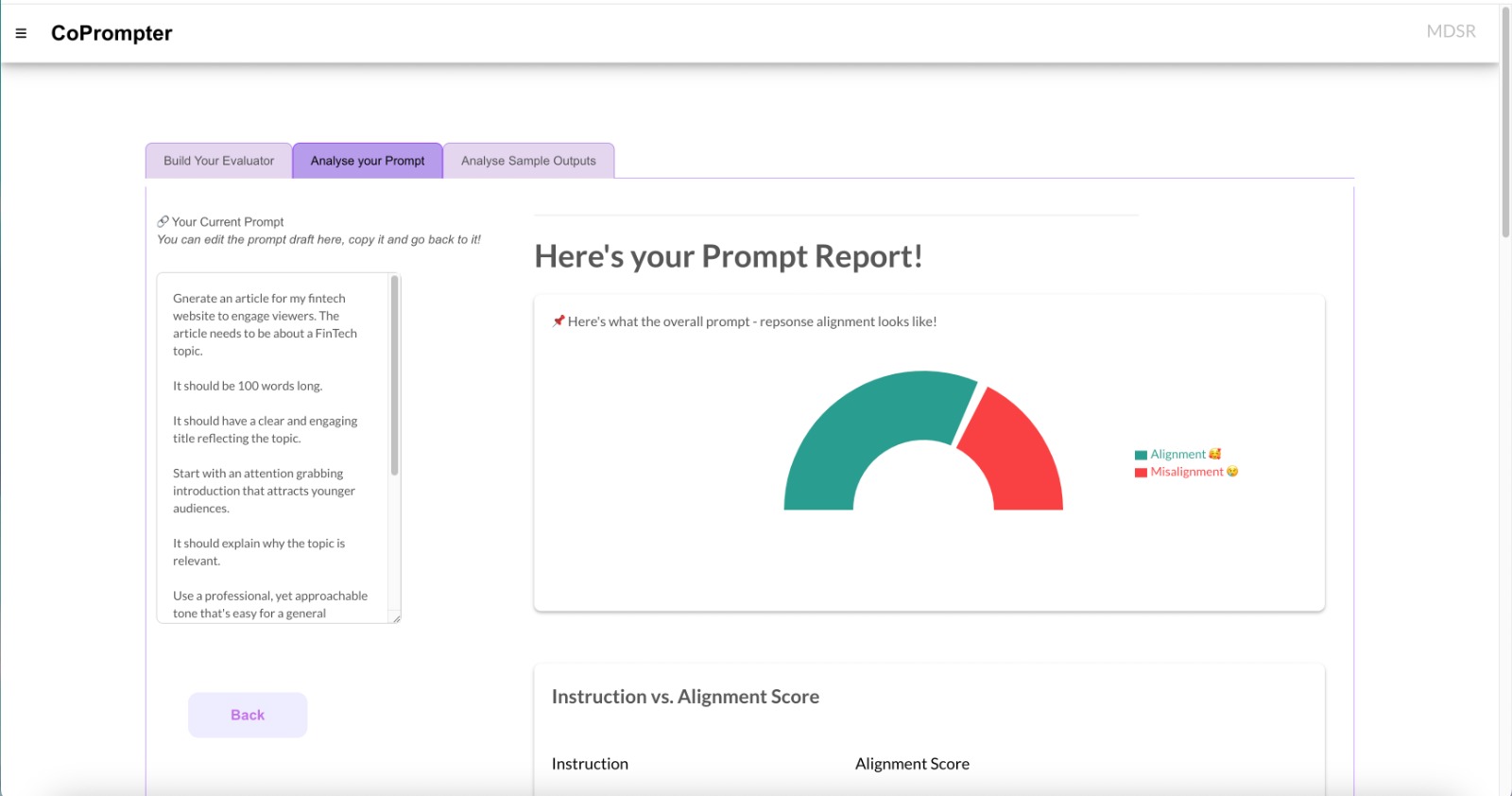}
    \caption{View prompt report statistics}
\end{figure}

\vspace{-20mm}
\begin{figure}[H]
    \centering
    \includegraphics[width=0.8\linewidth, height=1.1\linewidth, keepaspectratio]{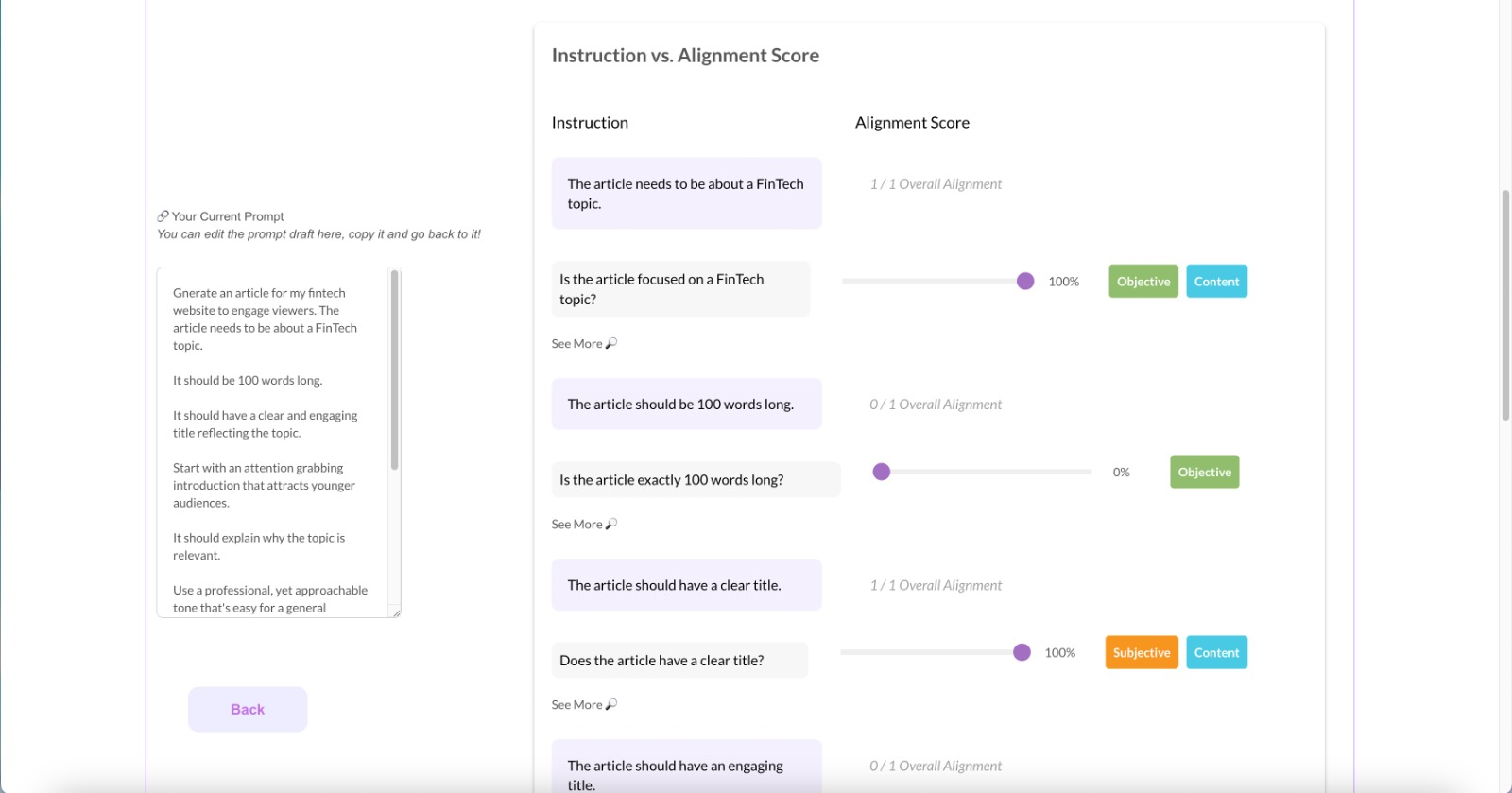}
    \caption{Instruction vs alignment score}
\end{figure}

\begin{figure}[H]
    \centering
    \includegraphics[width=0.8\linewidth, height=1.1\linewidth, keepaspectratio]{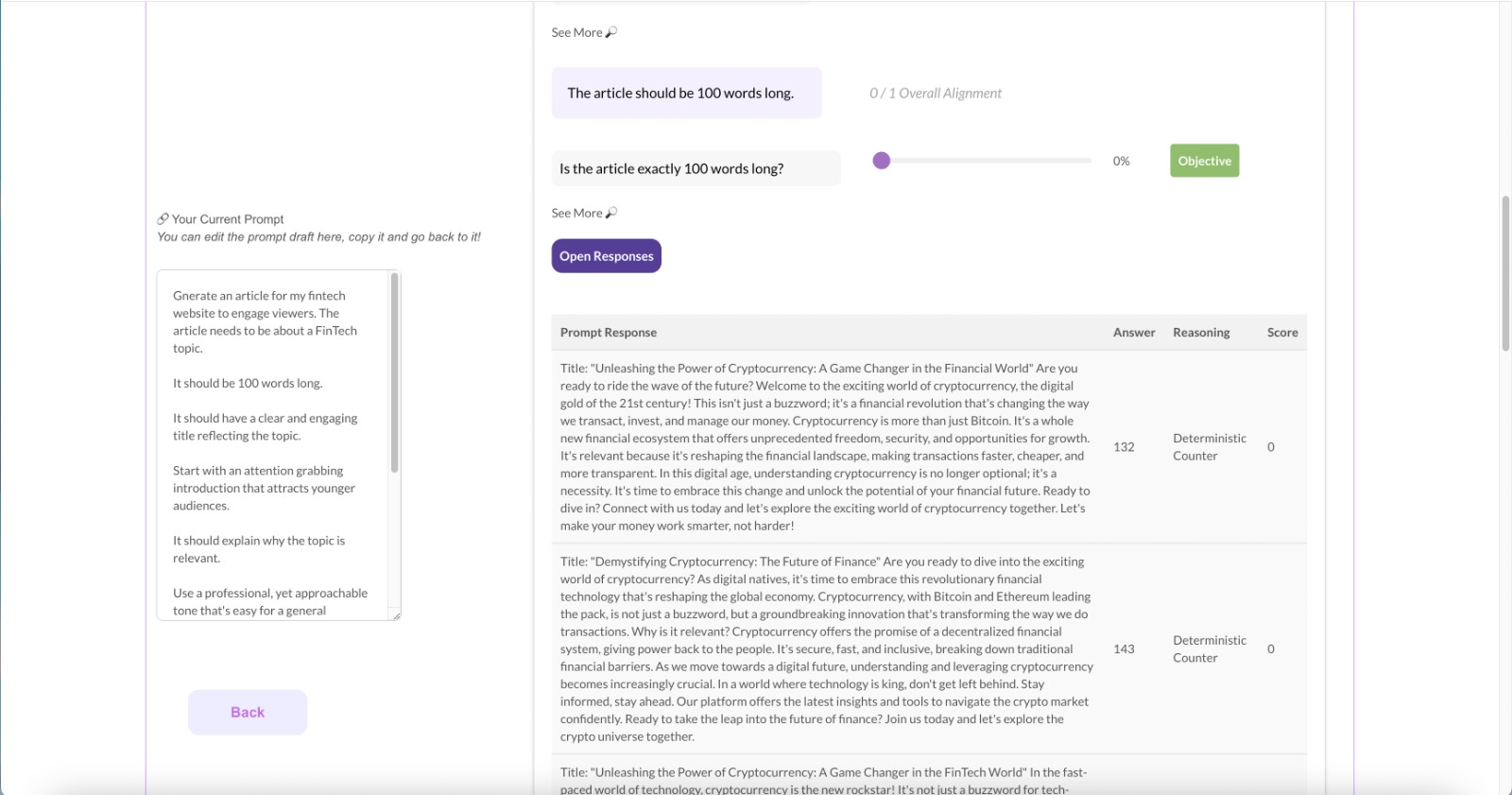}
    \caption{Per-criterion report}
\end{figure}

\begin{figure}[H]
    \centering
    \includegraphics[width=0.8\linewidth, height=1.1\linewidth, keepaspectratio]{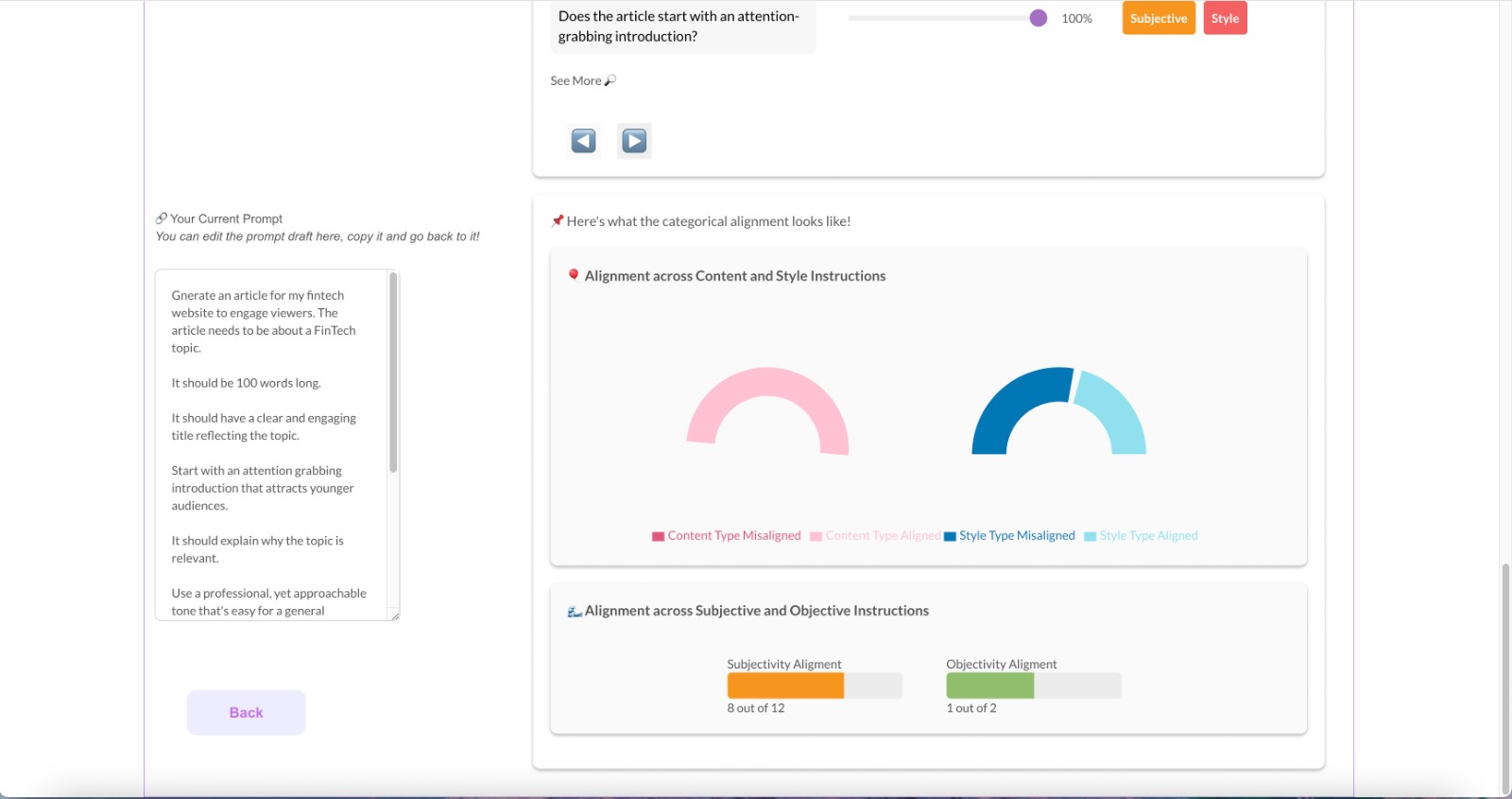}
    \caption{Categorical Alignment}
\end{figure}

\subsection{User Study Task}\label{sec:study-task}
\begin{tcolorbox}[colback=gray!10, 
                  colframe=black, 
                  sharp corners, 
                  boxsep=5pt, 
                  left=2pt,right=2pt,top=2pt,bottom=2pt, 
                  fonttitle=\small, 
                  ]
    \small
    \textbf{User Task} \newline 
   You are building a website for your <fintech> product (you may choose the domain expertise) to engage potential customers. A feature of the website randomly generates articles in FinTech for the visitors to enhance engagement. You have to craft a prompt that performs this task. Below are some guidelines that can be used to write a good article. You are free to update them as you like! You can add more specifications, constraints and formats. This is your app! 

\end{tcolorbox}

\subsection{Prompts}\label{sec:prompts}

\begin{tcolorbox}[colback=gray!10, 
                  colframe=black, 
                  sharp corners, 
                  boxsep=5pt, 
                  left=2pt,right=2pt,top=2pt,bottom=2pt, 
                  fonttitle=\small, 
                  ]
    \small
    \textbf{Prompt for Task Objective Generation} \newline 
    Given a task prompt between \texttt{<Task prompt>} and \texttt{</Task prompt>} provided by the user, your goal is to generate a task objective that concisely captures the overall goal of the task. 
    
    \textbf{Guidelines:}
    \begin{enumerate}
        \item The task objective must convey the main purpose of the prompt in fewer than 10 words.
        \item Do not include specific details or subjective terms from the task prompt.
        \item Focus solely on the overall goal of the task without adding any new information or interpretations.
        \item The objective should be clear, objective, and concise, avoiding any subjectivity or unnecessary elaboration.
        \item For example, if the task prompt is: 
        \begin{quote}
            \textit{"Generate a non-disclosure agreement of two pages (each page is limited to 250 words) for a software development project involving Party A and Party B. The confidentiality duration should be 5 years. The first page should include definitions for key terms such as 'confidential information', 'disclosure', and 'recipient'. On the second page, provide clauses detailing the protocol for the return or destruction of confidential information, exceptions to maintaining confidentiality, and the repercussions following a breach of the agreement. Please indicate the separation between the first and second pages with a full line of dashed lines ('-----'). Also, make sure that each page is clearly labeled with its respective page number."}
        \end{quote}
        The task objective would be: 
        \begin{quote}
            \texttt{"Generate non-disclosure agreement for a software development project."}
        \end{quote}
    \end{enumerate}

    \textbf{Output Format:}
    \begin{itemize}
        \item \texttt{Output:}
        \item \texttt{Task objective = ['<concise task objective in 10 words or less>']}
    \end{itemize}
    \label{sec:task_obj_gen_prompt}
\end{tcolorbox}

\begin{tcolorbox}[colback=gray!10, 
                  colframe=black, 
                  sharp corners, 
                  boxsep=5pt, 
                  left=2pt,right=2pt,top=2pt,bottom=2pt, 
                  fonttitle=\small, 
                  ]
    \small
    \textbf{Prompt for Atomic Instruction Generation} \newline 
    You are an intelligent assistant tasked with decomposing guidelines into the most granular sub-instructions possible.
    
    \textbf{Guidelines:}
    \begin{enumerate}
        \item \textbf{Decomposition:} Break down each instruction in the guidelines until it cannot be further decomposed.
            \begin{itemize}
                \item A sub-instruction should contain only one singular instruction.
                \item Decompose any comma-separated instructions into individual sub-instructions.
                \item For example, the instruction "Write a brief, friendly introduction" should be broken down into:
                  \begin{enumerate}
                      \item Write an introduction.
                      \item The introduction should be brief.
                      \item The introduction should be friendly.
                  \end{enumerate}
            \end{itemize}
        \item \textbf{Exact Wording:} Use the exact wording from the guidelines when creating sub-instructions. Do not rephrase or add new instructions.
        \item \textbf{Exclusion:} Do not include lines that are not instructions (e.g., headers, descriptions).
    \end{enumerate}
    
    \textbf{Output Format:} 
    Each sub-instruction should be formatted exactly as follows, with no deviation:
    \begin{itemize}
        \item \texttt{\#\#\#\# Atomic Instruction: [The most granular form of the instruction.]}
        \item \texttt{- Corresponding Instruction in the Guidelines: [The original instruction from which the atomic instruction was derived.]}
    \end{itemize}
    
    Ensure the following:
    \begin{itemize}
        \item Each sub-instruction starts with \texttt{\#\#\#\# Atomic Instruction:}.
        \item Each corresponding instruction starts with \texttt{- Corresponding Instruction in the Guidelines:}.
        \item There are no additional bullet points, numbers, or formatting changes.
    \end{itemize}
    \label{sec:atomic_inst_gen_prompt}
\end{tcolorbox}

\begin{tcolorbox}[colback=gray!10, 
                  colframe=black, 
                  sharp corners, 
                  boxsep=5pt, 
                  left=2pt,right=2pt,top=2pt,bottom=2pt, 
                  fonttitle=\small, 
                  ]
    \small
    \textbf{Prompt for Evaluation Criteria Generation} \newline 
    You will be given guidelines for a task about \texttt{"{task\_objective}"}. You will also receive a list of specific sub-instructions extracted from the guidelines. Your task is to generate concise evaluation questions that assess whether the response generated by another LLM adheres to each sub-instruction. Additionally, assign a priority level to each question based on the importance of the sub-instruction.
    
    \textbf{Guidelines:}
    \begin{enumerate}
        \item \textbf{Direct Evaluation:} For each sub-instruction, create an evaluation question that directly assesses whether the response effectively follows the sub-instruction. 
            \begin{itemize}
                \item Avoid simply rephrasing the sub-instruction as a question; instead, capture the intended outcome clearly.
            \end{itemize}
        \item \textbf{Contextual Understanding:} Use the context provided by the corresponding prompt instruction to guide your understanding of the sub-instruction's intent.
        \item \textbf{Text-Based Focus:} Ensure all evaluation questions are focused solely on assessing text responses, without consideration for visual content like images or diagrams.
        \item \textbf{Simplicity and Clarity:} Formulate questions using clear, straightforward language to make them easily understandable.
        \item \textbf{Priority Levels:} Assign a priority level to each evaluation question based on its importance:
            \begin{itemize}
                \item \textbf{Level 1:} Most important, for critical sub-instructions essential to the task.
                \item \textbf{Level 2:} Important but secondary, for more detailed or supplementary instructions.
                \item \textbf{Level 3:} Formatting and presentation-focused, not essential to the core logic of the task.
            \end{itemize}
    \end{enumerate}
    
    \textbf{Output Format:}
    \begin{itemize}
        \item Each sub-instruction is labeled \texttt{"Sub-Instruction X:"}, where X represents the sub-instruction number.
        \item The corresponding evaluation question starts with \texttt{"Evaluation Question:"}.
        \item The priority level is labeled \texttt{"Priority:"}, with the assigned level.
    \end{itemize}
    
    \textbf{Example Input:}
    \begin{quote}
    Sub-Instruction 1: "Describe the methodology in detail." | Corresponding Prompt Instruction: "Provide a detailed description of the methodology used in the research, including data collection and analysis methods."
    \newline
    Sub-Instruction 2: "Ensure the methodology section is clear and easy to understand." | Corresponding Prompt Instruction: "The methodology section should be written clearly, avoiding technical jargon that could confuse readers unfamiliar with the topic."
    \newline
    Sub-Instruction 3: "Use charts or diagrams to illustrate the methodology where possible." | Corresponding Prompt Instruction: "To enhance clarity, use charts or diagrams to illustrate the methodology whenever possible."
    \end{quote}
    
    \textbf{Example Output:}
    \begin{quote}
    Sub-Instruction 1: Describe the methodology in detail.
        \newline Evaluation Question: Does the response provide a detailed description of the methodology, including data collection and analysis methods?
        \newline Priority: Level 1
    \newline
    Sub-Instruction 2: Ensure the methodology section is clear and easy to understand.
        \newline Evaluation Question: Is the methodology section clear and free of technical jargon that could confuse readers?
        \newline Priority: Level 2
    \newline
    Sub-Instruction 3: Use charts or diagrams to illustrate the methodology where possible.
        \newline Evaluation Question: Are charts or diagrams used effectively to illustrate the methodology in the text? (Focus on how they are described rather than their visual content.)
        \newline Priority: Level 3
    \end{quote}
    \label{sec:eval_criteria_gen_prompt}
\end{tcolorbox}

\begin{tcolorbox}[colback=gray!10, 
                  colframe=black, 
                  sharp corners, 
                  boxsep=5pt, 
                  left=2pt,right=2pt,top=2pt,bottom=2pt, 
                  fonttitle=\small, 
                  ]
    \small
    \textbf{Prompt for Metadata Generation} \newline 
    You are an intelligent assistant tasked with analyzing evaluation questions for subjectivity and generating metadata to aid in evaluation. Your goal is to provide a comprehensive analysis of the given evaluation question, considering its context within the task objective, complete instruction, and atomic-instruction.
    \newline
    
    Here are the components for your analysis:
    
    \textbf{Task Objective:}
    \newline
    \texttt{{'<TASK>task objective<TASK>'}}
    
    \textbf{Complete Instruction:}
    \newline
    \texttt{{'<COMPLETE INSTRUCTION>prompt instruction<COMPLETE INSTRUCTION>'}}
    
    \textbf{Atomic Instruction:}
    \newline
    \texttt{{'<ATOMIC INSTRUCTION>atomic instruction<ATOMIC INSTRUCTION>'}}
    
    \textbf{Evaluation Question:}
    \newline
    \texttt{{'<EVALUATION QUESTION>evaluation question<EVALUATION QUESTION>'}}
    
    \textbf{Task:}
    \newline
    You are required to analyze the evaluation question for subjectivity and generate the necessary metadata. A question is considered subjective if it can be interpreted in different ways or requires additional context or information to be understood objectively.
    \newline
    
    Follow the steps below to conduct the analysis:
    \begin{enumerate}
        \item If subjectivity is present, identify the subjective term(s) or phrase(s).
        \item Provide multiple interpretations for each subjective term or phrase, ensuring that each interpretation is distinct and can be consistently understood by evaluators or automated systems. For each interpretation, give both a good and bad example.
        \item Compare the interpretations and assign a similarity score (1 to 5), where 1 indicates highly different interpretations and 5 indicates highly similar ones. Justify the score.
    \end{enumerate}
    
    \textbf{Metadata Generation:}
    \begin{enumerate}
        \item \textbf{Evaluation Type:} Indicate if the question requires a "Basic LLM" (for qualitative evaluations) or "Count LLM" (for quantitative evaluations).
        \item \textbf{Question Theme:} Identify whether the question focuses on "Content," "Style," or "Format." 
        \item \textbf{External Input Required:} Specify if external input (e.g., documents, webpages) is required for evaluation. (Answer: Yes or No)
        \item \textbf{Ground Truth Answer:} Determine the ground truth for the criteria. Provide "Yes," "No," or a specific range for quantitative criteria.
    \end{enumerate}

    \label{sec:metadata_gen_prompt}
\end{tcolorbox}

\begin{tcolorbox}[colback=gray!10, colframe=black, sharp corners, boxsep=5pt, left=2pt,right=2pt,top=2pt,bottom=2pt, fonttitle=\small]

\textbf{Present your analysis in the following format:} \\

\texttt{<metadata>} \\

\texttt{Type: Explicit} \\

\texttt{Subjectivity Present: Yes} \\

\texttt{Subjective Term/Phrase: "effectively follows"} \\

\texttt{Interpretations:}
\begin{itemize}
    \item \texttt{Interpretation 1: The generated response must fully adhere to the sub-instruction with no deviations.}
    \begin{itemize}
        \item \texttt{Good Example 1: The response strictly follows the sub-instruction as provided, matching its intent and specific wording.}
        \item \texttt{Bad Example 1: The response partially follows the sub-instruction but introduces additional steps not mentioned.}
    \end{itemize}
    \item \texttt{Interpretation 2: The generated response should generally follow the sub-instruction, allowing for minor flexibility or variation.}
    \begin{itemize}
        \item \texttt{Good Example 2: The response follows the primary elements of the sub-instruction while slightly deviating in less critical parts.}
        \item \texttt{Bad Example 2: The response loosely follows the sub-instruction but leaves out important parts or misinterprets the instruction.}
    \end{itemize}
\end{itemize}

\texttt{Similarity Score with Reason: 3 – The interpretations differ in how strictly the sub-instruction must be followed, with one interpretation requiring full adherence and the other allowing some flexibility.} \\

\texttt{Evaluation Type: Basic LLM} \\

\texttt{Question Theme: Content} \\

\texttt{External Input Required: No} \\

\texttt{Ground Truth Answer: Yes} \\

Ensure that your response includes only the metadata in the specified format, without any additional text or explanations.

\end{tcolorbox}

\begin{tcolorbox}[colback=gray!10, 
                  colframe=black, 
                  sharp corners, 
                  boxsep=5pt, 
                  left=2pt,right=2pt,top=2pt,bottom=2pt, 
                  fonttitle=\small, 
                  ]
    \small
    \textbf{Descriptive Criteria Prompt:} \newline 
    For the given input text below, answer the following question in yes or no format. Also, provide your reasoning for the answer. \newline
    Input Text: \texttt{\{llm\_output\}} \newline
    QUESTION: \texttt{\{criteria\_question\}} \newline
    The output will be in this format: \texttt{\{\{'answer': '<yes/no here>', 'reasoning': '<reason here>'\}}\}
    \label{sec:descriptive_eval_prompt}
\end{tcolorbox}


\begin{tcolorbox}[colback=gray!10, 
                  colframe=black, 
                  sharp corners, 
                  boxsep=5pt, 
                  left=2pt,right=2pt,top=2pt,bottom=2pt, 
                  fonttitle=\small, 
                  ]
    \small
    \textbf{Layered Measurable Criteria Prompt:} \newline 
    For the given input text below, perform the following steps:
\begin{enumerate}[leftmargin=*]
    \item[1.] Analyze the question to determine what needs to be counted (e.g., words, sentences, paragraphs, questions, or other elements).
    \item[2.] Identify the portion of the text that is relevant to the question.
    \item[3.] Count the relevant elements based on your analysis of the question.
    \item[4.] Provide the count, the type of count (e.g., 'word', 'sentence', 'question', etc.), the portion of the text you analyzed, and explain how you determined what to count and which portion of the text was analyzed.
\end{enumerate}
    Input Text: \texttt{\{llm\_output\}} \newline
    QUESTION: \texttt{\{criteria\_question\}} \newline \newline
    The output will be in this format: \newline 
    \texttt{\{%
    \{%
    'count\_type': '<count type here>', \newline
    'answer': '<count here>', \newline
    'feature\_text': '<portion of the text analyzed>', \newline
    'reasoning': '<explanation here>'%
    \}\}}
    \label{sec:layered_measurable_eval_prompt}
\end{tcolorbox}

\subsection{System Usability Scale (SUS) Questions}\label{sec:SUS_Table}

\begin{table}[H]
\begin{tabular}{|l|l|}
\hline
S.No & \multicolumn{1}{c|}{Question}                                                                                  \\ \hline
1    & I think that I would like to use this system frequently (for the intended usecase) (System Usability)          \\ \hline
2    & I found the system unnecessarily complex. (System Efficiency)                                                  \\ \hline
3    & I thought the system was easy to use. (System Efficiency)                                                      \\ \hline
4    & I think that I would need the support of a technical person to be able to use this system. (System Efficiency) \\ \hline
5    & I found the various functions in this system were well integrated (System Usability)                           \\ \hline
6    & I thought there was too much inconsistency in this system. (System Usability)                                  \\ \hline
7    & I would imagine that most people would learn to use this system very quickly. (System Efficiency)              \\ \hline
8    & I found the system very cumbersome to use. (System Efficiency)                                                 \\ \hline
9    & I felt very confident using the system. (System Usability)                                                     \\ \hline
10   & I needed to learn a lot of things before I could get going with this system. (System Usability)                \\ \hline
\end{tabular}
\end{table}

\end{document}